\documentclass[%        	Class options:
aps,%                   	American Physical Society
prd,%                   	Physical Review D
superscriptaddress,%    	Authors' addresses linked with superscripts
nofootinbib,%           	Does not treat footnotes as references
floatfix]%              	Fixes float errors
{revtex4}%              	REVTEX 4 Package used	
\usepackage{graphicx,%  	Default Latex 2eps package for embedding figures
longtable,%             	Useful for long table
amsmath,%					AMS math 
color,%						Colours
bm}%						Bold math symbols
\begin{document}
%======================================================================================
\title{Mapping reactor neutrino spectra from TAO to JUNO}
%--------------------------------------------------------------------------------------
%
\author{        	Francesco~Capozzi}
\affiliation{   	Max-Planck-Institut f\"ur Physik (Werner-Heisenberg-Institut), 
					F\"ohringer Ring 6, 80805 M\"unchen, Germany}
\author{        	Eligio~Lisi}
\affiliation{   	Istituto Nazionale di Fisica Nucleare, Sezione di Bari, %\\
               		Via Orabona 4, 70126 Bari, Italy}
\author{        	Antonio~Marrone}
\affiliation{   	Dipartimento Interateneo di Fisica ``Michelangelo Merlin,'' %\\
               		Via Amendola 173, 70126 Bari, Italy}%
\affiliation{   	Istituto Nazionale di Fisica Nucleare, Sezione di Bari, %\\
               		Via Orabona 4, 70126 Bari, Italy}
\begin{abstract}%.......................................................................
The Jiangmen Underground Neutrino Observatory (JUNO) project aims at probing, at the same time, 
the two main frequencies of three-flavor neutrino oscillations, as well as their interference
related to the mass ordering (normal or inverted), at a distance of $\sim 53$ km from two powerful reactor
complexes in China, at Yangjiang and Taishan. In the latter complex, the unoscillated spectrum from one
reactor core is planned to be closely monitored by
the Taishan Antineutrino Observatory (TAO), expected to have   
better resolution ($\times 1/2$) and higher statistics ($\times 30$) than JUNO. In the 
context of $\nu$ energy spectra endowed with fine-structure features from summation calculations,  
we analyze in detail the effects of energy resolution and nucleon recoil on observable event spectra. 
We show that a model spectrum in TAO can be mapped into a corresponding spectrum in JUNO through 
appropriate convolutions. The mapping is exact in the hypothetical case without oscillations,
and holds to a very good accuracy in the real case with oscillations.  
We then analyze the sensitivity to mass ordering of JUNO (and its precision oscillometry capabilities) 
assuming a single reference spectrum, as well as bundles of variant spectra,
as obtained by changing nuclear input uncertainties in summation calculations from a publicly 
available toolkit. We show through a $\chi^2$ analysis that variant spectra induce little reduction of the sensitivity
in JUNO, especially when TAO constraints are included. Subtle aspects of the statistical analysis
of variant spectra are also discussed.    
\end{abstract}%.........................................................................
\maketitle

%%%%%%%%%%%%%%%%%%%%%%%%%%%%%%%%%%%%%%%%%%%%%%%
\section{Introduction}
\label{Sec:Intro}
%%%%%%%%%%%%%%%%%%%%%%%%%%%%%%%%%%%%%%%%%%%%%%%

Experiments based on electron antineutrinos ($\overline\nu_e$) from nuclear reactors---referred to 
as reactor neutrinos hereafter---have marked the history of neutrino physics 
\cite{History,Bilenky:2019qcm,Jarlskog:2019axp,Vogel:2019fnm,Lasserre:2019spd}. In the neutrino oscillation 
era they have been---and continue to be---a major tool for both discoveries and precision measurements 
\cite{Cao:2017drk,Qian:2018wid,Antonelli:2020uui}. In particular, reactor
experiments at long baseline (LBL) \cite{Gando:2013nba} and short baseline (SBL)
\cite{Adey:2018zwh,Bak:2018ydk,DoubleChooz:2019qbj} have 
observed the oscillation patterns governed by the mass-mixing parameters $(\Delta m^2_{21},\, \theta_{12})$, 
and $(\Delta m^2_{32},\,\theta_{13})$, respectively \cite{PDG}. 
At medium baselines (MBL), reactor experiments with high statistics and
resolution could observe both patterns and their interference,  
probing  $\alpha =\mathrm{sign}(\Delta m^2_{32}/\Delta m^2_{21})=\pm 1$ and thus the $\nu$ mass ordering, either ``normal'' 
(NO, $\alpha=+1)$ or ``inverted'' (IO, $\alpha=-1$) \cite{Petcov:2001sy}. In order to perform such MBL oscillation searches, 
as well as a wider
physics program, the Jiangmen Underground Neutrino Observatory (JUNO) is being built near Kaiping (China), at equal baselines ($L\sim 53$~km) 
from the Taishan and Yangjiang reactor complexes \cite{Li:2013zyd,An:2015jdp}.

In this context, the neutrino energy spectra at the reactor source(s) represent important inputs,
that should be understood and computed with an accuracy 
comparable  to the experimental one. Reactor neutrino spectra have usually been obtained 
by either conversion from measured electron spectra (``conversion''  approach)
or by summing over thousands of beta transitions tabulated in nuclear databases (``ab initio'' or ``summation'' approach),
and sometimes in combination (``hybrid'' approach) \cite{Hayes:2016qnu,Huber:2016fkt}. 
In the last decade, these approaches have been challenged by new data (or by reanalyses of old data) 
that do not compare well with computed spectra, even invoking nonstandard physics such as sterile neutrinos 
\cite{Dentler:2017tkw,Giunti:2019qlt,Berryman:2020agd} (not considered herein). A primary example 
is the unexpected ``bump'' observed around 5 MeV in current SBL oscillation experiments 
\cite{Adey:2018zwh,Bak:2018ydk,DoubleChooz:2019qbj,Ko:2016owz} 
(and possibly in older data  \cite{Zacek:2018bij}), whose 
understanding is still entangled with many issues, including: 
normalization anomalies in the total flux and its fuel components  
\cite{Huber:2016xis,Gebre:2017vmm,Giunti:2017nww,Hayes:2017res,Adey:2018qct,Adey:2019ywk,RENO:2018pwo,Ashenfelter:2018jrx}, 
incomplete information in nuclear databases \cite{Hayes:2015yka,Sonzogni:2016yac,Sonzogni:2017wxy,Ma:2018fqf}, 
possible energy-scale systematics \cite{Mention:2017dyq},
suppression of $\beta$-decay spectra systematics  \cite{Hardy:1977suw} via total absorption 
\cite{Fallot:2012jv,Rasco:2016leq,Guadilla:2019gws,Guadilla:2019zwz,Guadilla:2019aiq,Estienne:2019ujo}
and other techniques \cite{Guadilla:2020pjj},
and, on the theory side, improved 
calculations of (allowed and forbidden) $\beta$ decay spectra  
\cite{Hayes:2013wra,Wang:2017htp,Li:2019quv,Fang:2015cma,Fang:2020emq,Yoshida:2018zga,Petkovic:2019wyw,Hayen:2018uyg,Hayen:2019ieh}.

Another layer of complexity, pointed out in summation calculations of the neutrino energy spectrum, is the presence of sawtooth-like 
substructures, as expected from Coulomb effects in individual $\beta$ decays 
\cite{Dwyer:2014eka,Sonzogni:2015aoa}. These fine-structure features have not been observed within the resolution of current experiments, with the possible exception of a hint discussed in 
\cite{Sonzogni:2017voo}. Observing or constraining (at least a few) prominent substructures
in future, high-resolution and high-statistics reactor neutrino experiments, would be beneficial both for
nuclear spectroscopy (allowing to pin down the spectral contributions of specific 
fission products \cite{Sonzogni:2017voo}), and for neutrino oscillometry
(reducing small-scale fuzzy uncertainties that might affect the JUNO sensitivity to mass ordering 
\cite{Forero:2017vrg}). Although the latter benefit may be marginal if one assumes 
``known'' substructures from nuclear databases \cite{Danielson:2018tzi,Cheng:2020ivh}, the observation of unexpected 
spectral anomalies at large energy scales (normalization and bump issues) provides a warning about the possible emergence of
``unknown'' features also at small scales. For a recent and comprehensive overview of current issues and future prospects 
in understanding reactor neutrino spectra, see the contributions in \cite{IAEA-Vienna}.

In oscillation searches at reactor experiments, spectral uncertainties can be efficiently suppressed 
by comparing near (unoscillated) and far (oscillated) event spectra 
\cite{Mikaelyan:1998yg,Mikaelyan:1999pm}, as performed in \cite{Adey:2018zwh,Bak:2018ydk,DoubleChooz:2019qbj}. 
In the context of JUNO, a concept for a high-resolution near detector was mentioned in \cite{Wang-2017} and further detailed 
in \cite{Zhan-2018,Wonsak-2018}. 
This concept has evolved into a full-fledged project, the Taishan Antineutrino Observatory (TAO)
\cite{Cao-2019,Ranucci:2018erv,Wang-2019,Sisti-2019,Smirnov-2020}.%
%--------------------------------------------------------------------------
\footnote{While this paper was being written, the complete TAO Conceptual Design Report (CDR) was released \cite{TAO-CDR}. 
For the purposes of our work, the CDR confirms the basic characteristics of TAO that we have adopted from previous 
reports \cite{Cao-2019,Ranucci:2018erv,Wang-2019,Sisti-2019,Smirnov-2020}.}
%--------------
TAO is expected to monitor the
unoscillated spectrum emitted
by one of the Taishan nuclear reactors, with a gain of about $\times 1/2$ in energy
resolution and $\times 30$ in event statistics 
with respect to the oscillated spectrum at JUNO.
Independently of neutrino oscillations, 
high-resolution spectral measurements at TAO will set unprecedented benchmarks \cite{Cao-2019} 
for research in nuclear fission physics \cite{Brown:2018jhj,Schmidt:2018hwz,Bernstein:2019nqq} and for broader investigations of the  
neutrino-nuclear response in particle physics and astrophysics \cite{Ejiri:2019ezh}. In general, progress in 
neutrino and nuclear physics, coupled with precision measurements at TAO, is expected to 
significantly constrain the range of neutrino spectral models to be used in future JUNO data analyses.

In this work we build upon our previous studies of precision oscillometry \cite{Capozzi:2013psa,Capozzi:2015bpa}, but considering
summation spectra with substructures and possible uncertainties.
We use the publicly available toolkit Oklo \cite{Oklo,Littlejohn:2018hqm} to generate ensambles of spectra within
quoted errors on yields, branching ratios and endpoint energies for each decay.  This toolkit, although 
currently not updated in terms of nuclear database inputs (taken as of 2015 \cite{Oklo}), is appropriate 
for our methodological purposes and numerical experiments. For simplicity, we shall assume 
that the underlying neutrino spectra are the same in TAO and JUNO.
In reality, the former will closely monitor only one reactor core
in Taishan, while the latter will detect a signal generated by several reactors in both Taishan and Yangjiang,
with different fuel evolutions \cite{Cao-2019,Ciuffoli:2019nli,TAO-CDR}. The related fuel   
corrections will require detailed information and modeling for each reactor, that are 
beyond the scope of this paper and will be studied elsewhere.

Our work can be divided in two main parts. 
In the first part (Secs.~\ref{Sec:Unosc} and \ref{Sec:Osc}) we discuss 
the formal relations between the TAO and JUNO energy spectra.  
We start by revisiting in detail the effects of
resolution and recoil that, although well known in principle, are not always properly
distinguished and implemented at the level of accuracy required by future measurements.  
Then we show that any observable energy spectrum of events in TAO can be mapped into a
corresponding spectrum in JUNO by a proper convolution. In particular, we show that
this  mapping can be exactly performed in the hypothetical case of no oscillations (Sec.~\ref{Sec:Unosc})
and can be very 
accurately generalized, via an ansatz, to the real case  of neutrino oscillations (Sec.~\ref{Sec:Osc}).
These results allow to predict the JUNO spectrum directly from a model for the observable 
event spectrum at the TAO detector, rather than from a model for the unobservable neutrino spectrum at the reactor source. 

In the second part of the paper we perform quantitative studies of the mass-ordering sensitivity and 
precision oscillometry in JUNO, first by considering a single reference spectrum 
with substructures, and then by adding bundles of variant spectra to be constrained by TAO.
In Sec.~\ref{Sec:Single} we revisit our previous
analysis \cite{Capozzi:2015bpa} including the reference Oklo spectrum, new priors for
the oscillation parameters, and reduced error bands for  smooth flux-shape and energy-scale systematics.
We confirm the accuracy of the mapping and discuss the impact of these new inputs. 
In Sec.~\ref{Sec:Ensemble} we generate bundles of Oklo variants around the previous
reference spectrum. We perform a $\chi^2$ analysis of variant spectra in JUNO, alternative to
the Fourier analysis in \cite{Danielson:2018tzi}, and highlight several statistical
issues arising from sampling the nuclear input uncertainties. By varying all the known nuclear data inputs,
we generate and analyze an ensemble of $10^5$ spectra in JUNO, but find
no reduction of the sensitivity to mass ordering, with or without TAO; we trace 
this unexpected result to subtle undersampling issues in the generated bundle of spectra. 
We repeat the analysis by constructing an equally
numerous but ``more densely sampled'' bundle, and find a small reduction 
of the JUNO sensitivity, consistent with \cite{Danielson:2018tzi} 
and improved with the help of TAO. 
These results, based on ``known'' nuclear
inputs,  suggest some cautionary  comments on 
parametrizations of ``unknown'' substructure uncertainties,
as those considered in \cite{Forero:2017vrg}.   
We also analyze the JUNO accuracy on the relevant mass-mixing parameters, 
which is found to be basically unaffected by fine-structure issues.
Our results are summarized in Sec.~\ref{Sec:Final}. 

%%%%%%%%%%%%%%%%%%%%%%%%%%%%%%%%%%%%%%%%%%%%%%%
\section{Mapping the spectrum from TAO to JUNO without oscillations}
\label{Sec:Unosc}
%%%%%%%%%%%%%%%%%%%%%%%%%%%%%%%%%%%%%%%%%%%%%%%

Reactor neutrinos can be detected through the inverse beta decay (IBD) process $\overline\nu_e+p\to e^+ + n$ followed by $e^+$
annihilation  and delayed $n$ capture.  
Using the notation of \cite{Capozzi:2013psa,Capozzi:2015bpa}, we
focus on two energy variables for IBD events:
%............................................................................
\begin{eqnarray}
E &=& \mathrm{unobservable}\ \overline\nu_e\ \mathrm{energy},\\
E_\mathrm{vis} &=& \mathrm{observable\ (visible)\ energy\ of\ the\ event}.  
\end{eqnarray}
%............................................................................
We also consider the unobservable neutrino spectrum $S_\nu$, as given by the reactor $\nu$ flux
 $\Phi(E)$ times the IBD cross section
\cite{Vogel:1999zy,Strumia:2003zx} $\sigma_\nu(E)$,
%............................................................................
\begin{equation}
\label{Nuspec}
S_\nu(E) = \Phi(E)\sigma_\nu(E),
\end{equation}
%............................................................................
and the observable IBD event spectrum at the detector $X$,
%............................................................................
\begin{equation}
S_X = S_X(E_\mathrm{vis}),\ X=T,\,J,
\end{equation}
%............................................................................
where the subscripts $T$ and $J$ shall refer to TAO and JUNO, respectively.

In this Section we show that, for no oscillation, the TAO spectrum can be exactly mapped into the JUNO spectrum,
%............................................................................
\begin{equation}
S_T(E_\mathrm{vis}) \to S_J(E_\mathrm{vis}), 
\end{equation}
%............................................................................
without knowing {\em a priori\/} $S_\nu(E)$. This result is nontrivial in the presence of resolution and 
recoil effects, that we discuss below following \cite{Capozzi:2013psa}. The mapping will be extended to the
oscillation case in the next Section~\ref{Sec:Osc}.

%%%%%%%%%%%%%%%%%%%%%%%%%%%%%%%%%%%%%%%%%%
\subsection{Detector resolution}

In a detector with perfect energy resolution, $E_\mathrm{vis}$ would be equal to %............................................................................
\begin{equation}
\label{Nores}
E_\mathrm{vis} = E_e + m_e \ \mathrm{(perfect\ resolution)},
\end{equation} 
%............................................................................
where  $E_e$ and $m_e$ are the total $e^+$ energy and mass, respectively.

In reality, due to finite photon statistics and other instrumental effects, 
$E_\mathrm{vis}$ is distributed around $E_e+m_e$ according to a resolution function $r_X$,     
%............................................................................
\begin{equation}
\label{Gaussian}
r_X(E_\mathrm{vis},\,E_e+m_e\,|\,\sigma^2_X) = \frac{1}{\sqrt{2\pi\sigma^2_X}} \exp
\left( {-\frac{1}{2}\frac{(E_\mathrm{vis}-E_e-m_e)^2}{\sigma^2_X}}\right), 
\end{equation} 
%............................................................................
where $\sigma^2_X$ is the energy resolution variance for the detector $X$. 
For TAO we adopt, as  a representative value of $\sigma_T$   
\cite{Cao-2019,Ranucci:2018erv,Wang-2019,Sisti-2019,Smirnov-2020},
%............................................................................
\begin{equation}
\label{TAOres}
\frac{\sigma_T}{E_\mathrm{vis}} = \frac{1.7\%}{\sqrt{E_\mathrm{vis}}}\ ,
\end{equation}
%............................................................................
while for JUNO we take $\sigma_J/E_\mathrm{vis}$ as in \cite{Capozzi:2015bpa} (roughly equal to $3\%/\sqrt{E_\mathrm{vis}}$).

\subsection{Nucleon recoil}

If recoil effects were neglected in IBD events, $E_e+m_e$ would be related to $E$ via
%............................................................................
\begin{equation}
\label{Norec}
E_e + m_e = E - 0.783\ \mathrm{MeV}\ \mathrm{(no\ recoil)}. 
\end{equation} 
%............................................................................
Nucleon recoil induces an angle-dependent deficit in $E_e$, making this relation an upper bound. In general, 
$E_e$ ranges between two kinematical extrema $E_{1,2}(E)$ 
\cite{Strumia:2003zx},
%............................................................................
\begin{equation}
\label{Interval}
E_1 \leq E_e \leq  E_2\  (< E - m_e - 0.783\ \mathrm{MeV})
\end{equation} 
%............................................................................
with a relatively flat distribution (see \cite{Capozzi:2013psa} and Fig.~2 
therein).
As in \cite{Capozzi:2013psa,Capozzi:2015bpa} we approximate
this distribution through a top-hat function, 
%............................................................................
\begin{equation}
\label{Tophat}
t(E,\,E_e)=\frac{1}{\sigma_\nu(E)}\frac{d\sigma_\nu(E,\,E_e)}{dE_e}\simeq 
\left\{
\begin{array}{cl}
(E_2-E_1)^{-1} & \mathrm{for\ } E_1\leq E_e\leq E_2,\\
0              & \mathrm{otherwise},
\end{array}
\right.
\end{equation}
%............................................................................
where $d\sigma_\nu/dE_e$ is the differential IBD cross
section \cite{Strumia:2003zx}. We have explicitly checked for TAO 
(as we did in \cite{Capozzi:2013psa} for JUNO) that corrections to this approximation, 
named hereafter as ``full recoil'', are numerically irrelevant in spectral calculations (not shown).

We also consider a less accurate approximation,
dubbed as ``mid-recoil,'' whereby  
 the midpoint of the interval  in Eq.~(\ref{Interval})
is taken as a proxy for $E_e$ \cite{Vissani:2014doa},
%............................................................................
\begin{equation}
\label{Midrec}
E_e\simeq E_e^\mathrm{mid} = (E_1+E_2)/2,
\end{equation}  
%............................................................................
and the Jacobian 
%............................................................................
\begin{equation}
J(E)=(dE_{e}^\mathrm{mid}/dE)^{-1} 
\end{equation}  
%............................................................................
is included, when passing
from neutrino to positron energy spectra, to ensure event number conservation.
A useful approximation for $E_e^\mathrm{mid}$ (and thus for $J$) is given in \cite{Vissani:2014doa} as
%............................................................................
\begin{eqnarray}
E_{e}^\mathrm{mid} (E) &\simeq& \frac{E-\Delta_E}{1+\frac{E}{m_p}},\\
J(E) &\simeq & \frac{\left(1+\frac{E}{m_p}\right)^2}{1+\frac{\Delta_E}{m_p}},
\end{eqnarray}
%............................................................................
where $\Delta_E=m_e+0.783$~MeV. 
This mid-recoil recipe captures well 
the average recoil shift but ignores its energy spread, which is definitely nonnegligible in TAO 
as shown below. 

%%%%%%%%%%%%%%%%%%%%%%%%%%%%%%%%%%%%%%%%%%%%%%%%%%%%%%%%%%%%%%%%%%%%%%%%%%%%%%%%%%%%%%%%%%
\begin{figure}[t]
\begin{minipage}[c]{.95\textwidth}
\includegraphics[width=.95\textwidth]{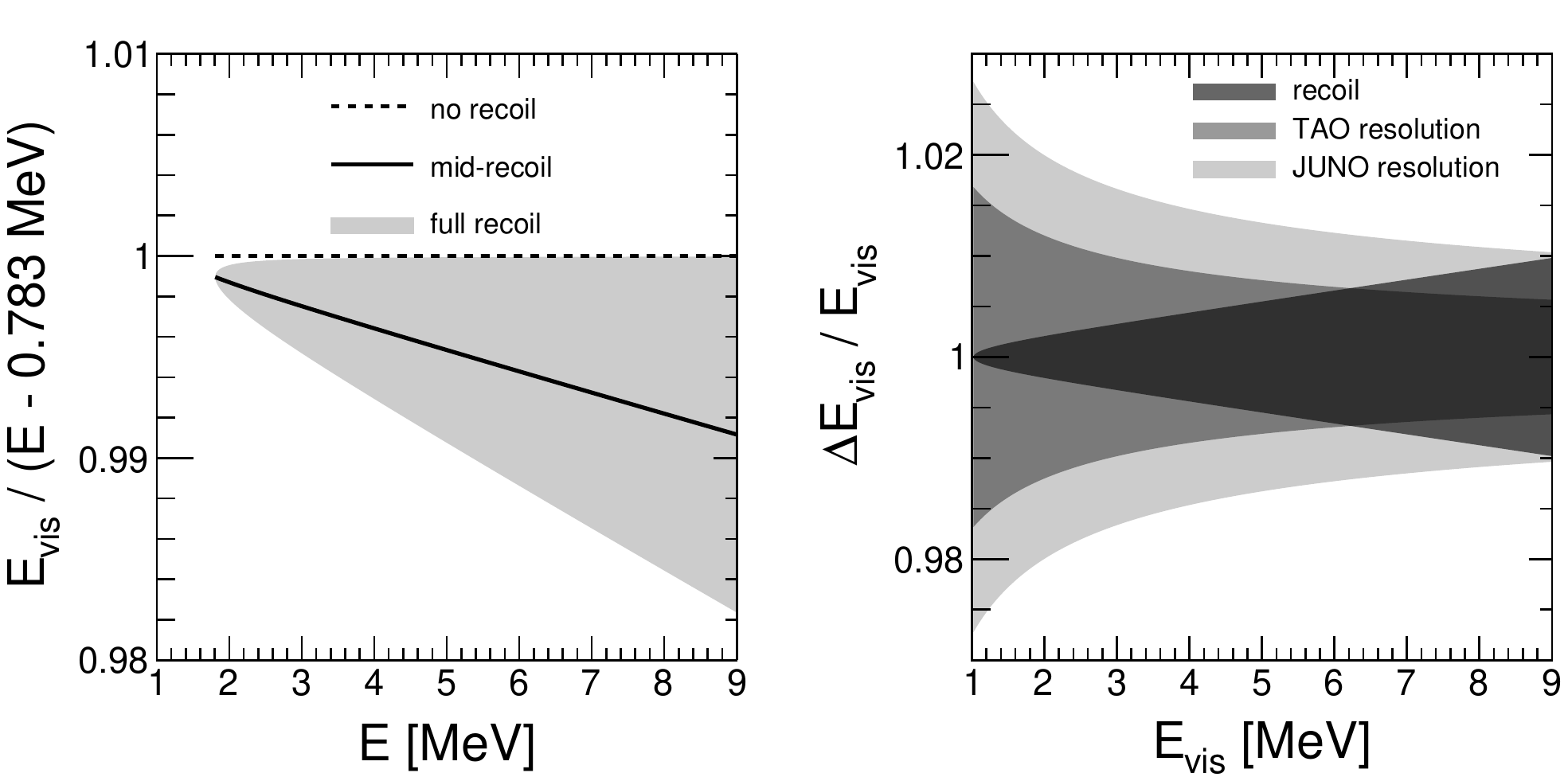}
\caption{\label{Fig_01}
\footnotesize 
Left panel: fractional recoil effects in terms of neutrino energy. Right panel: energy spread due to recoil, compared
with $1\sigma$ resolution widths in TAO and JUNO. In both panels, the recoil band is bounded by kinematical limits.
See the text for details.} 
\end{minipage}
\end{figure}
%%%%%%%%%%%%%%%%%%%%%%%%%%%%%%%%%%%%%%%%%%%%%%%%%%%%%%%%%%%%%%%%%%%%%%%%%%%%%%%%%%%%%%%%%%

\subsection{Resolution and recoil effects: comparison and combination}

If both resolution and recoil effects were neglected, then Eqs.~(\ref{Nores}) and (\ref{Norec})
would lead to the often-quoted approximation
$ E_\mathrm{vis} = E-0.783~\mathrm{MeV}$.
Figure~\ref{Fig_01} (left panel) shows the recoil corrections to such relation
as a function of neutrino energy $E$, in terms
of deviations from unity [dashed line at $1\equiv E_\mathrm{vis}/(E-0.783~\mathrm{MeV})$].  
The gray area corresponds to the the one-sided energy deficit due to full recoil effects
[Eq.~(\ref{Interval})], while the solid line marks the mid-recoil approximation [Eq.~(\ref{Midrec})].        
Notice that, at high reactor neutrino energies, the visible event energy is both
shifted and smeared out at the percent level.
In Fig.~\ref{Fig_01} (right panel) we show the fractional energy spread 
$\Delta E_\mathrm{vis}/E_\mathrm{vis}$ due to recoil and resolution, separately.
In particular, $\Delta E_\mathrm{vis}$ is shown as
$\pm (E_2-E_1)/2$ for recoil (dark gray),  as
$ \pm \sigma_T$ for  TAO (gray band) and as $\pm \sigma_J$ for JUNO (light gray).
Recoil and resolution effects in TAO appear to be of comparable size,
and none of them can be neglected in accurate spectral analyses, especially in view of their
impact on the observability of substructures.

As shown in \cite{Capozzi:2013psa}, the combination of the resolution and recoil effects 
is fully encoded
in an energy resolution function $R_X$ that connects the relevant energies 
$E_\mathrm{vis}$ and $E$, as obtained by convolving the gaussian distribution $r_X$ 
in Eq.~(\ref{Gaussian}) with the top-hat distribution $t$ in Eq.~(\ref{Tophat}),
%............................................................................
\begin{equation}
\begin{aligned}
\label{Erf}
R_X(E_\mathrm{vis},\, E\,|\,\sigma^2_X) &= r_X \ast t \\
&=
\frac{1}{2(E_2-E_1)}
\left[
\mathrm{erf}\left(
\frac{E_\mathrm{vis}-(E_1+m_e)}{\sqrt{2\sigma^2_X}}
\right)-
\mathrm{erf}
\left(
\frac{E_\mathrm{vis}-(E_2+m_e)}{\sqrt{2\sigma^2_X}}
\right) 
\right]
\end{aligned}
\end{equation}
%............................................................................
where $\sigma_X=\sigma_X(E_\mathrm{vis})$, while
the dependence on $E$ comes from $E_{1,2}=E_{1,2}(E)$ that we take from the full calculation in \cite{Strumia:2003zx}; 
 see \cite{Capozzi:2013psa} for further details, including the adopted convention for the error function (erf).

The observable energy spectra $S_X$ of IBD events in TAO and JUNO (in the absence of oscillations) can then be 
computed by convolving the neutrino spectrum $S_\nu$ in Eq.~(\ref{Nuspec}) with the above resolution function,
%............................................................................
\begin{equation}
\begin{aligned}
S_X(E_\mathrm{vis}) &= {\cal N}_X \ S_\nu \ast R_X \\
&= {\cal N}_X \int_{E_T}^{\infty} dE \ S_\nu(E) \ R_X(E_\mathrm{vis},\, E\,|\,\sigma^2_X),\   X=T,\,J\ ,
\end{aligned}
\end{equation}
%............................................................................
where $E_T=1.806$~MeV is the IBD $\nu$ energy threshold and 
${\cal N}_X$ is a normalization factor.

%%%%%%%%%%%%%%%%%%%%%%%%%%%%%%%%%%%%%%%%%%%%%%%%%%%%%%%%%%%%%%%%%%%%%%%%%%%%%%%%%%%%%%%%%%
\begin{figure}[t]
\begin{minipage}[c]{0.92\textwidth}
\includegraphics[width=.92\textwidth]{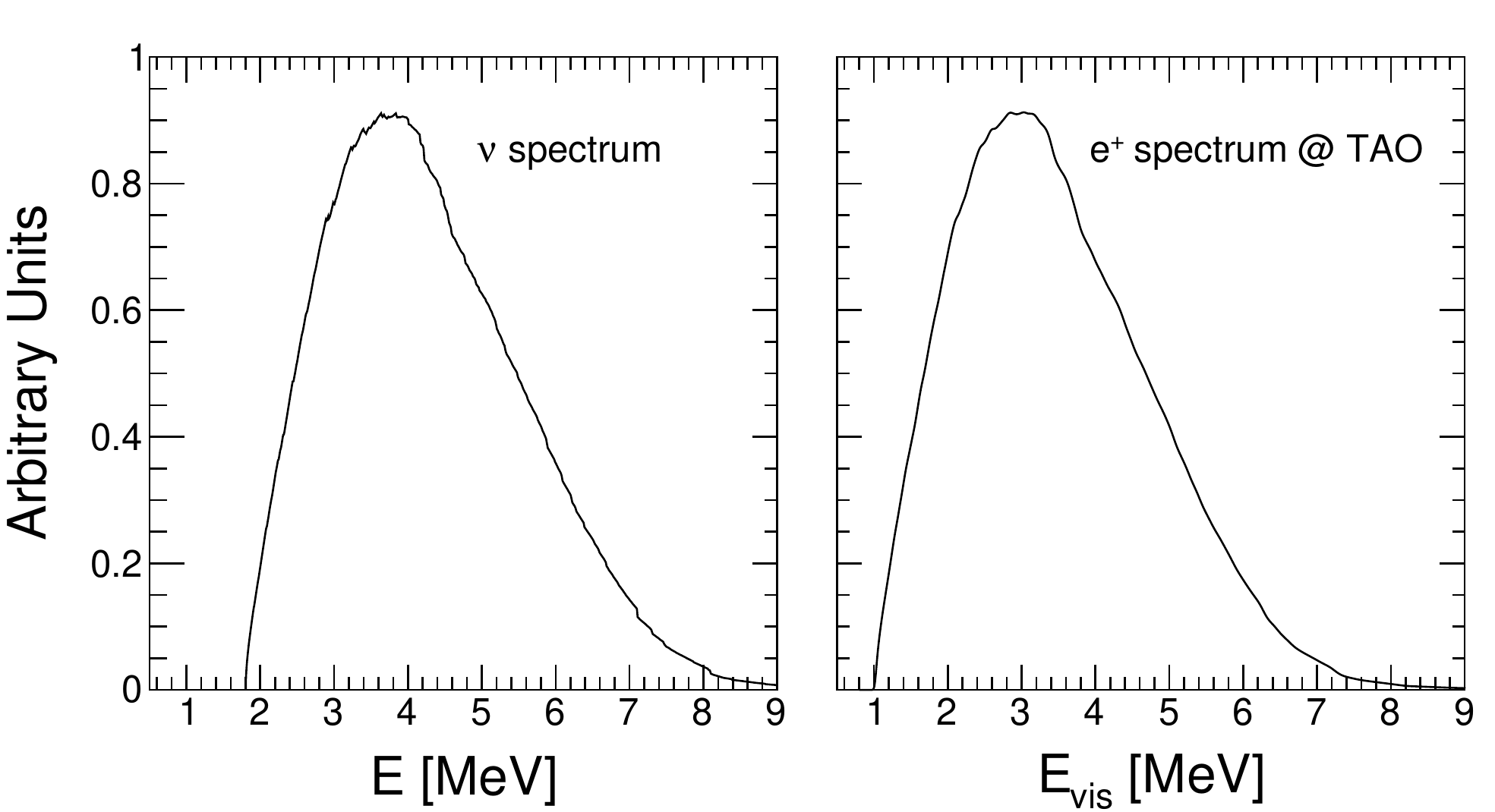} 
\caption{\label{Fig_02}
\footnotesize 
Reference neutrino spectrum as obtained from the Oklo toolkit \cite{Oklo} (left panel)
and corresponding visible energy spectrum at TAO, including recoil and resolution effects (right panel).
For graphical comparison, the two spectra are normalized
to the same area, in arbitrary units. 
} 
\end{minipage}
\end{figure}
%%%%%%%%%%%%%%%%%%%%%%%%%%%%%%%%%%%%%%%%%%%%%%%%%%%%%%%%%%%%%%%%%%%%%%%%%%%%%%%%%%%%%%%%%%
 
Figure~\ref{Fig_02} shows in the left panel the neutrino energy spectrum  $S_\nu$, 
as obtained with default nuclear input parameters for the $\nu$ flux $\Phi$ from the
Oklo toolkit \cite{Oklo}, times the cross section from \cite{Strumia:2003zx}.
 The TAO visible energy spectrum $S_T$ is shown in the right panel, including recoil and energy resolution effects. Spectra are normalized to the same area (in arbitrary units) to facilitate comparison in shape. It can be seen that spectral substructures in $S_\nu$ (sawtooth and step-like features)
are smeared out in $S_T$, but still partly visible. Such substructures would no longer be visible in JUNO (not shown). 

%%%%%%%%%%%%%%%%%%%%%%%%%%%%%%%%%%%%%%%%%%%%%%%%%%%%%%%%%%%%%%%%%%%%%%%%%%%%%%%%%%%%%%%%%%
\begin{figure}[t]
\begin{minipage}[c]{0.95\textwidth}
\includegraphics[width=.95\textwidth]{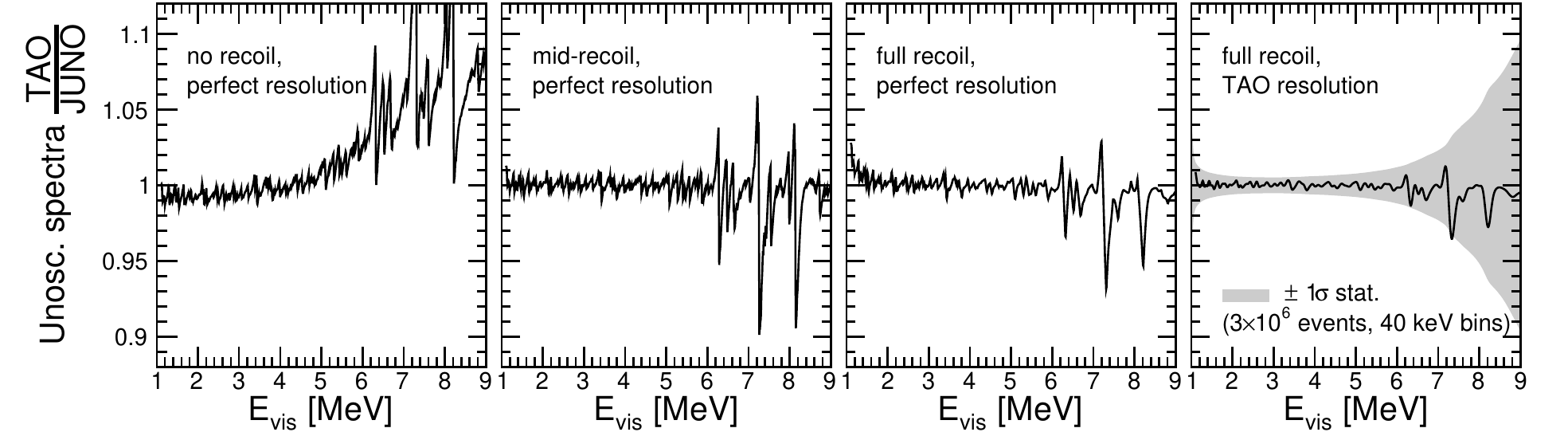}
\caption{\label{Fig_03}
\footnotesize 
Ratio $S_T/S_J$ of visible energy spectra in TAO and JUNO (unoscillated), normalized to the same area for
comparison. The denominator $S_J$ always include full recoil and resolution effects. The
numerator $S_T$ includes recoil and resolution effects in progression from left to right. In the rightmost plot, the spectral ratio 
substructures are compared with the $\pm 1$ statistical error band in TAO, assuming $3\times 10^6$ events and 40~keV 
bin width.
} 
\end{minipage}
\end{figure}
%%%%%%%%%%%%%%%%%%%%%%%%%%%%%%%%%%%%%%%%%%%%%%%%%%%%%%%%%%%%%%%%%%%%%%%%%%%%%%%%%%%%%%%%%%

Figure~\ref{Fig_03} shows the ratio $S_T/S_J$ of the unoscillated energy spectra in TAO and JUNO (arbitrarily normalized to the
same area). From left to right, the numerator $S_T$ is calculated with progressive inclusion of recoil and resolution effects, 
while the denominator $S_J$ always includes all such effects. In particular, the first three panels assume perfect energy resolution 
in TAO ($\sigma_T=0$), with increasingly accurate treatments for nucleon recoil. In the first plot (no recoil) the spectral ratio
shows evident substructures and a high-energy excess (spectral tilt), due to neglected energy recoil losses that also bias the substructure peak positions  by up to 1\% (not visible by eye). 
 In the second plot (mid-recoil
approximation including the Jacobian) the average energy losses are accounted for, the shift disappears,  and
the subtructures are correctly aligned in energy.  
In the third plot (full recoil treatment) the inclusion of the recoil energy spread suppresses the finest spectral structures 
and, at high energy, reduces their amplitudes by a factor of $\sim2$. Finally, further suppression of fine structure features (and 
another amplitude reduction by a factor of $\sim2$ or more) is due to the inclusion of the finite TAO 
resolution width $\sigma_T$ from Eq.~(\ref{TAOres}) in the rightmost panel. In this panel we also show the $\pm 1\sigma$ error band
in TAO assuming $3\times 10^6$ events, i.e., $\sim 30$ times the statistics expected in JUNO \cite{Cao-2019} 
in the presence of oscillations for about 5 years (that amounts to $\sim 100,000$ events \cite{Capozzi:2013psa}). 
The statistical band depends on the bin width, here taken as 40~keV (25 bins per MeV interval) in order to cover 
the most prominent substructures within a few bins at least. It can be seen that a handful of fine-structure
features reaches the $\sim 1\sigma$ level in amplitude, allowing TAO to probe spectral models with
different predicted substructures (see also \cite{Cao-2019}).
We shall discuss some statistical issues concerning the model selectivity of TAO in Sec.~\ref{Sec:Ensemble}.

Summarizing, resolution and recoil effects in the TAO energy spectrum are of comparable size and should be carefully
implemented, in order to avoid energy biases and unrealistic amplitudes for fine-structure spectral features.   
Resolution effects produce a gaussian smearing (whose width decreases with increasing energy), while recoil
effects produce an energy shift plus a top-hat smearing 
(whose width increases with increasing energy).  
Their combination (convolution)
leads to an analytical expression for the energy resolution function \cite{Capozzi:2013psa} as in Eq.~(\ref{Erf}),
that can be usefully applied to the calculation of both TAO and JUNO spectra.

Finally we mention that, in principle, the impact of recoil effects may be reduced by directional information in the final state 
of IBD events, see \cite{Wei:2020yfs} for a recent proposal in the context of TAO. We do not explore this option hereafter,
but surmise that constraining
recoil effects amounts to replace the function $t$ in Eq.~(\ref{Tophat}) with another one ($t'$) having
smaller variance, possibly leading to an analytical result as in Eq.~(\ref{Erf}) if the parameterization of $t'$ is simple.

\subsection{Mapping the spectrum from near to far}

The resolution function $R_J$ in Eq.~(\ref{Erf}) for JUNO is obtained by convolving a gaussian $r_J$ 
having a variance $\sigma^2_J$  
with a top-hat function $t$. In turn, $r_J$ can be thought as the convolution of two gaussians $r_T$ and $r_D$ with variances given,
respectively, by $\sigma^2_T$ (as in TAO) and by 
%............................................................................
\begin{equation}
\sigma^2_D(E_\mathrm{vis})=\sigma^2_J(E_\mathrm{vis})-\sigma^2_{T}(E_\mathrm{vis})>0,
\end{equation}
%............................................................................
that is, the difference between the energy resolution variances in JUNO and TAO.

Then, through convolutions, one gets an exact mapping from TAO to JUNO (unoscillated) spectra as follows:
%............................................................................
\begin{equation}
\begin{aligned}
\label{Mapping}
S_J(E_\mathrm{vis})	& =  S_\nu \ast R_J \\
					& =  S_\nu \ast r_J\ast t\\
					& =  S_\nu \ast r_D\ast r_T \ast t\\
					& =  S_J \ast r_D\\
					& =  \int_0^\infty dE_\mathrm{vis}' \ S_J(E_\mathrm{vis}') 
					\ r_D(E_\mathrm{vis},\,E_\mathrm{vis}'\,|\,\sigma^2_D)\ ,    
\end{aligned}
\end{equation}
%............................................................................
where  normalization factors ${\cal N}_X$
have been dropped for simplicity, and $r$ has the same functional form as in Eq.~(\ref{Gaussian}), with $E_e+m_e$ replaced by
$E_\mathrm{vis}'$.

This analytical result has a simple physical interpretation: The JUNO unoscillated spectrum in visible energy ($S_J$) can be obtained from the TAO spectrum ($S_T$) by applying an extra gaussian smearing with variance $\sigma^2_D$, equal to the difference of variances in
JUNO ($\sigma^2_J$) and TAO ($\sigma^2_T$). In doing so, recoil effects remain correctly implemented in both TAO and JUNO. 

Note that Eq.~(\ref{Mapping}) directly relates the
observable event spectra $S_T$ and $S_J$, without using the unobservable neutrino spectrum $S_\nu$. 
This represents an advantage in terms
of nuclear physics modeling: Constructing
a model for $S_T$ (compatible with future TAO data) will generally be less demanding than 
building a complete model for $S_\nu$, since the former will exhibit
only a few surviving substructures to be properly described via summation.

A final comment is in order. As stated in Sec.~\ref{Sec:Intro},  we are assuming the the TAO and JUNO spectra
are generated by the same underlying $\nu$ spectrum $S_\nu$. 
However, JUNO will collect a neutrino flux also from reactor cores different from the one monitored by TAO, 
leading to fuel-component differences in the reference $S_\nu$ and 
to corrections to the ideal case in Eq.~(\ref{Mapping}). Fuel evolution issues and related spectral effects in TAO versus JUNO 
are beyond the scope of this investigation, and will be treated in a future work; 
see \cite{Cao-2019,Ciuffoli:2019nli,TAO-CDR} for useful considerations in this context.

%%%%%%%%%%%%%%%%%%%%%%%%%%%%%%%%%%%%%%%%%%%%%%%
\section{Mapping the spectrum from TAO to JUNO with oscillations}
\label{Sec:Osc}
%%%%%%%%%%%%%%%%%%%%%%%%%%%%%%%%%%%%%%%%%%%%%%%

In this Section we generalize the TAO~$\to$~JUNO mapping of Eq.~(\ref{Mapping}) in the presence 
of oscillations, characterized by a $\overline\nu_e$ survival probability $P_{ee}(E)$. An obstacle to this goal
is that the integrand $S_\nu(E)$ gets replaced by the product $S_\nu \cdot P_{ee}$ in JUNO,
and that the convolution of a product is not the product of convolutions, as also noted in \cite{Ciuffoli:2019nli}.

However, after reviewing the functional form of $P_{ee}$, we propose an ansatz that, to a very good approximation, overcomes 
this problem. We shall generalize Eq.~(\ref{Mapping}) by including an effective probability $P^\mathrm{eff}_{ee}$, 
expressed in terms of observable spectra $S_X$ and visible energy $E_\mathrm{vis}$, that
bypasses any prior knowledge of the (unobservable) neutrino energy spectrum $S_\nu(E)$. 
We shall then discuss the validity of this ansatz, and use it in an updated analysis of the JUNO
sensitivity to the neutrino mass ordering and to precision oscillometry.

\subsection{Oscillation probability in terms of neutrino energy}

In this subsection we describe the survival probability $P_{ee}(E)$, 
largely following \cite{Capozzi:2013psa} to which we refer the reader for details and references. 
In general, $P_{ee}(E)$ in JUNO depends on several parameters,
%............................................................................
\begin{equation}
\label{Param}
P_{ee}(E)=P_{ee}\left(E\,|\,\delta m^2,\,\Delta m^2,\alpha,\,\theta_{12},\, \theta_{13},\,N_e,\,\{w_n,\,L_n\}\right)
\end{equation}
%............................................................................
where $\delta m^2=m^2_2-m^2_1$ and $\Delta m^2=|m^2_3-(m^2_1+m^2_2)/2|>0$ are the squared mass splitting parameters, 
$\alpha=\pm 1$ distinguishes the mass ordering (normal or inverted), $\theta_{12}$ and $\theta_{13}$ are the 
mixing angles, $N_e$ is the electron density in matter, and $\{w_n,\,L_n\}$ characterizes the set of reactors, each  
contributing to the total flux with fractional weight $w_n$ ($\sum_n w_n=1$) at distance $L_n$, under the assumption of 
identical fuel components.

Useful derived parameters are
%............................................................................
\begin{equation}
\Delta m^2_{ee} = \Delta m^2 + \frac{\alpha}{2}(c^2_{12}-s^2_{12})\delta m^2,
\end{equation}
%............................................................................
where $c_{12}=\cos \theta_{12}$ and $s_{12}=\sin \theta_{12}$, and
%............................................................................
\begin{equation}
 \delta = \frac{\delta m^2 L}{4 E},\   \Delta_{ee} = \frac{ \Delta m^2_{ee}  L}{4 E}\ ,
\end{equation}
%............................................................................
where $L = \sum_n w_n L_n$ is the average baseline.
%............................................................................
%\begin{equation}
%L = \sum_n w_n L_n .
%\end{equation} 
%............................................................................
Matter effects in JUNO depend on the ratio
%............................................................................
%\begin{equation}
$\mu = ({2\sqrt{2}G_F N_e E})/{\delta m^2}$
%\end{equation}
%............................................................................
and lead to an effective mass-mixing  pair $(\tilde \delta,\,\tilde \theta_{12} )$  given by 
%............................................................................
%\begin{eqnarray}
$\tilde \delta \simeq \delta(1+\mu\cos2\theta_{12})$ and 
$\sin 2 \tilde \theta_{12} \simeq  \sin 2\theta_{12}(1-\mu \cos 2\theta_{12})$ 
%\end{eqnarray}
%............................................................................
at first order in the small parameter $\mu$ \cite{Capozzi:2013psa} (see also \cite{Li:2016txk,Khan:2019doq}).
 The dependence of $P_{ee}$ on its parameters can then be simply expressed as
%............................................................................
\begin{equation}
\label{Pmat}
P_{ee}(E) = c^4_{13}\tilde P + s^4_{13} + 2 s^2_{13}c^2_{13} 
\sqrt{\tilde P}\;w\;\cos(2\Delta_{ee}+\alpha\varphi),
\end{equation} 
%............................................................................
where 
%............................................................................
\begin{equation}
\label{Ptilde}
\tilde P = 1-4\tilde s^2_{12}\tilde c^2_{12}\sin^2\tilde \delta 
\end{equation}
%............................................................................
encodes $(\tilde \delta,\,\tilde \theta_{12})$ matter effects, while
%............................................................................
%\begin{equation}
$w \simeq 1-2\Delta_{ee}^2\sum_n w_n (1-L_n/L)^2$
%\end{equation} 
%............................................................................
is a damping factor due to the spread of baselines $L_n$, 
and $\varphi$ is the interference phase directly related to mass ordering \cite{Minakata:2007tn}.
An accurate empirical parameterization for $\varphi$ is given by \cite{Capozzi:2013psa}%
%----------------------------------------------------------------------------------
\footnote{Here we report a typo in Eq.~(45) of  \cite{Capozzi:2013psa}, 
where $\sin\delta$ should be replaced by $\sin 2\delta$. We thank A.~Formozov for detecting the misprint.}
%............................................................................
\begin{equation}
\varphi \simeq 2 s^2_{12}\delta \left(1-\frac{\sin 2\delta}{2\delta\sqrt{P} }\right),
\end{equation}
%............................................................................
where $P$ reads as in Eq.~(\ref{Ptilde}) but with vacuum mass-mixing values ($\delta,\,\theta_{12}$).%
%---------------------------------------------------------------------------------------
\footnote{Replacing ($\delta,\,P$) with ($\tilde\delta,\,\tilde P$)
in $\varphi$ leads to insignificant corrections to $P_{ee}$ \cite{Capozzi:2013psa}.}
%-----------

For the oscillation parameters in $P_{ee}$ [Eq.~(\ref{Pmat})] we assume the following priors (central values and $\pm1\sigma$,
after symmetrizing errors and averaging NO-IO differences)
from the global analysis in \cite{Capozzi:2018ubv}:
%............................................................................
\begin{eqnarray}
s^2_{12} &=& (3.04 \pm 0.13)\times 10^{-1},\label{prior1}\\
\delta m^2 &=& (7.34 \pm 0.16)\times 10^{-5}\ \mathrm{eV}^2,\label{prior2}\\
s^2_{13} &=& (2.16 \pm 0.08)\times 10^{-2}, \label{prior3}\\
\Delta m^2_{ee} &=& (2.448\pm 0.034)\times 10^{-3}\ \mathrm{eV}^2\label{prior4}.
\end{eqnarray}
%............................................................................
Determining the mass ordering in JUNO amounts to prove that in $P_{ee}$, 
besides the oscillation phase $2\Delta_{ee}$, there is and extra
an interference phase $\varphi$ endowed with a definite sign ($\alpha=\pm 1$)
and not scaling as $1/E$, otherwise it would be absorbed into a shift of $\Delta m^2_{ee}$
\cite{Parke:2016joa}; equivalently, one should find evidence for a non-constant ratio $\varphi/2\Delta_{ee}$.
It has been pointed out \cite{Qian:2012xh} that energy calibration errors at (sub)percent level may (partly) mimic 
$\varphi/2\Delta_{ee}\neq$~{\em const}
\cite{Li:2013zyd,Capozzi:2013psa,Capozzi:2015bpa}; 
in this context, future evidence for some substructures emerging in TAO spectrum, located
at the energies predicted by nuclear summation models, may help the overall calibration of the reference 
spectrum to be projected from TAO to JUNO (provided that JUNO is also accurately calibrated in energy). 
Correct implementation of recoil effects, in both TAO and JUNO, remains mandatory to avoid energy biases at comparable (sub)percent
levels.

\begin{figure}[t]
\begin{minipage}[c]{0.9\textwidth}
\includegraphics[width=.85\textwidth]{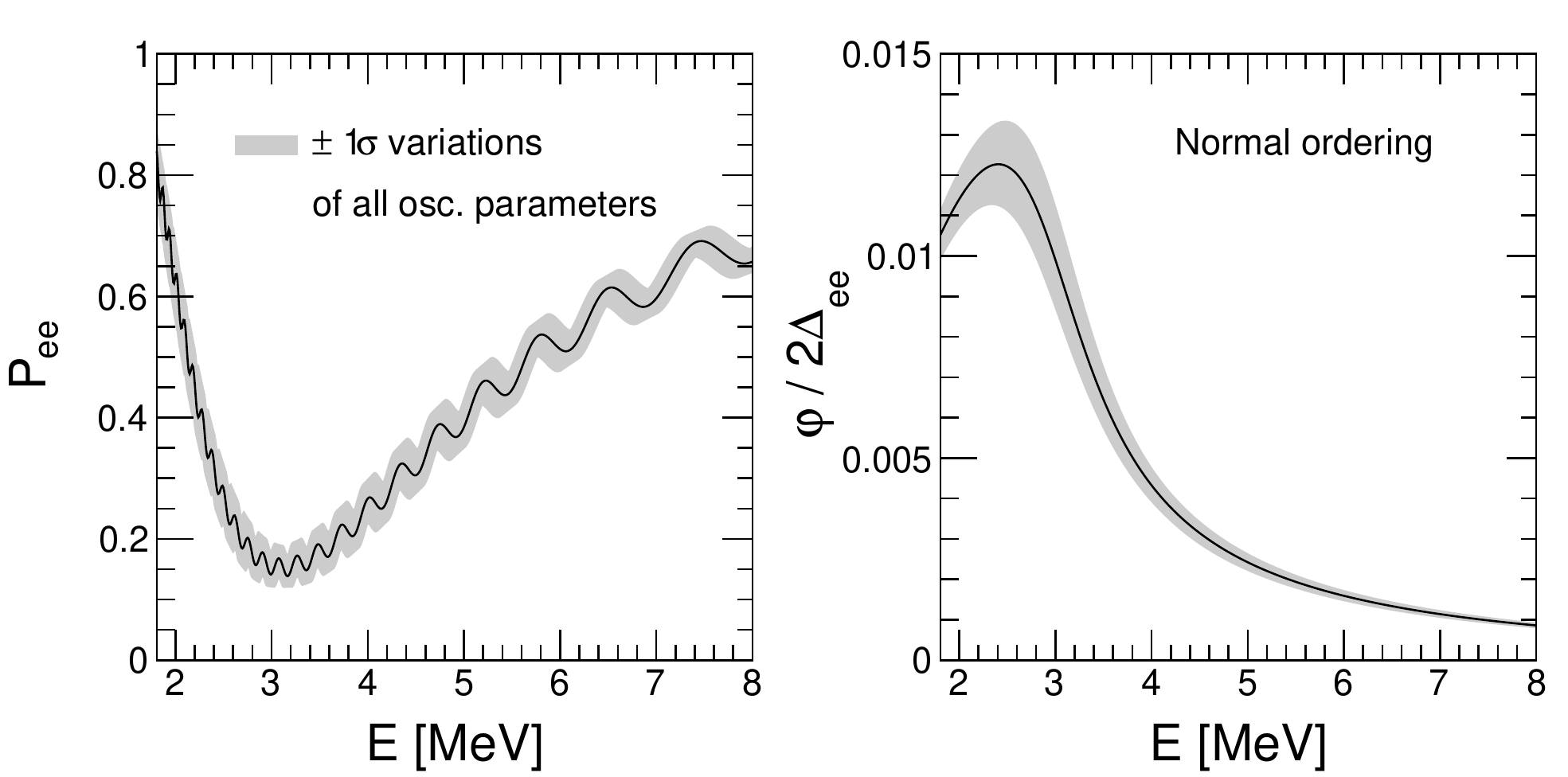}
\caption{\label{Fig_04}
\footnotesize 
Survival probability $P_{ee}$ (left panel) and oscillation phase ratio $\varphi/2\Delta_{ee}$ (right panel)
for electron antineutrinos with energy $E$ in JUNO. Solid lines are computed for central values of the
oscillation parameters, while the gray bands correspond to the envelope of $\leq 1\sigma$ variations 
in the prior ranges (see the text). Normal ordering is assumed. For inverted ordering, $P_{ee}$ would be similar
while $\varphi/2\Delta_{ee}$ would reverse its sign (not shown).} 
\end{minipage}
\end{figure}

Figure~\ref{Fig_04} shows the function $P_{ee}$ (left panel) and the ratio  $\varphi/2\Delta_{ee}$ (right panel) as a function
of energy. The solid curves and gray bands correspond, respectively, to the central values 
and to the envelopes of $\leq 1\sigma$ variations for the
oscillation parameters. Normal ordering is assumed. 
The smallness of $\varphi/2\Delta_{ee}$ 
illustrates the challenges of mass ordering determination at MBL reactors. Note the relatively high values of 
$\varphi/2\Delta_{ee}$ for $E\sim 3$~MeV may fractionally change by up to $\pm 8\%$ within the gray band, 
and that for similar energies $P_{ee}$ (and thus the IBD event rate) may also change by up to $\pm 12\%$. Therefore, 
variations of the oscillation parameters within their current global-fit errors can appreciably affect  the 
prospective mass ordering sensitivity in JUNO, as discussed later.

%%%%%%%%%%%%%%%%%%%%%%%%%%%%%%%%%%%%%%%%%%%%%%%%%%%%%%%%%%%%%%%%%%%%%%%%%%%%%%%%%%%%%%%%%%
%%%%%%%%%%%%%%%%%%%%%%%%%%%%%%%%%%%%%%%%%%%%%%%%%%%%%%%%%%%%%%%%%%%%%%%%%%%%%%%%%%%%%%%%%%

\subsection{Ansatz: effective probability in terms of visible energy}

Given the probability $P_{ee}(E)$, the oscillated spectrum at JUNO (including resolution
and recoil effects, and up to a normalization factor) is
%............................................................................
\begin{equation}
\label{ExactJ}
S_J(E_\mathrm{vis}) = \int_{E_T}^{\infty} dE \ S_\nu(E) \ P_{ee}(E) \ R_J(E_\mathrm{vis},\, E\,|\,\sigma^2_J).\   
\end{equation}
%............................................................................
Our goal is to obtain such $S_J$ by mapping the TAO spectrum $S_T$, in a form analogous to Eq.~(\ref{Mapping}), 
%............................................................................
\begin{equation}
\label{Mapping1}
S_J(E_\mathrm{vis}) =  \int_0^\infty dE_\mathrm{vis}' \ S_T(E_\mathrm{vis}')\ P^\mathrm{eff}_{ee}(E'_\mathrm{vis})
					\ r_D(E_\mathrm{vis},\,E_\mathrm{vis}'\,|\,\sigma^2_D),   
\end{equation} 
%............................................................................
where $P^\mathrm{eff}_{ee}$ should act as an effective oscillation probability, expressed in terms of the measured visible energy
rather than the unobservable neutrino energy. 
This problem is {\em exactly\/} solved by imposing, in the kernel of Eq.~(\ref{Mapping1}), that
%............................................................................
\begin{equation}
S_T(E'_\mathrm{vis})\ P_{ee}^\mathrm{eff} (E'_\mathrm{vis})=\int_{E_T}^\infty dE\ S_\nu(E)\ P_{ee}(E)\ R_T(E'_\mathrm{vis},\, E\,|\,\sigma^2_T)\ ,
\end{equation}
%............................................................................
namely, by defining $P^\mathrm{eff}_{ee}$ as follows
(with a change $E'_\mathrm{vis}\to E_\mathrm{vis}$ in the dummy variable):
%............................................................................
\begin{equation}
\label{Exact}
P_{ee}^\mathrm{eff}(E_\mathrm{vis})=\frac
{\int_{E_T}^\infty dE\ S_\nu(E)\ P_{ee}(E)\ R_T(E_\mathrm{vis},\, E\,|\,\sigma^2_T)}
{\int_{E_T}^\infty dE\ S_\nu(E)\ R_T(E_\mathrm{vis},\, E\,|\,\sigma^2_T)}\ ,
\end{equation}
%............................................................................
which represents the weighted average of $P_{ee}$ over the neutrino spectrum ($S_\nu$) times the TAO energy resolution function 
($R_T$). In a sense, $P_{ee}^\mathrm{eff}$ is a  smeared version of $P_{ee}$, averaged over
$S_\nu$ variations on an energy scale set by $\sigma_T$. 

However, this formally exact solution is not satisfactory, as it  requires the knowledge of the unobservable neutrino spectrum $S_\nu$. 
We make then the following ansatz, that replaces the unobservable 
$S_\nu$ with its closest observable proxy, namely $S_T$: Within the integral kernels of Eq.~(\ref{Exact}), 
the function $S\nu(E)$ is substituted by the
TAO spectral function $S_T(E_\mathrm{vis})$, and in turn $E_\mathrm{vis}$ is identified with its 
closest proxy $E_\mathrm{vis}(E)=E_e^\mathrm{mid}(E)+m_e$. 
Conservation of number of events is ensured by imposing $S_\nu(E)dE=S_T(E_e^\mathrm{mid}+m_e)dE_e^\mathrm{mid}$, so that
the complete replacement involves $J^{-1}(E)=dE_e^{\mathrm{mid}}/dE$: 
%............................................................................
\begin{equation}
S_\nu(E) \to S_T\left(E_e^\mathrm{mid}(E)+m_e)\right)J^{-1}(E).
\end{equation}
%............................................................................
The effect of the Jacobian in the above formula is rather small numerically, since $J(E)$ changes slowly with $E$
(if it were constant, it would be canceled in the ratio); we keep it for the sake of completeness.

Summarizing, our ansatz for the mapping $S_T \to S_J$ (including oscillations) consists in calculating 
an effective JUNO spectrum $S^\mathrm{eff}_J(E_\mathrm{vis})$ from the observable TAO spectrum $S_T(E_\mathrm{vis})$ as  
%............................................................................
\begin{equation}
\label{Mapping2}
S^\mathrm{eff}_J(E_\mathrm{vis}) =  \int_0^\infty dE_\mathrm{vis}' \ S_T(E_\mathrm{vis}')\ P^\mathrm{eff}_{ee}(E'_\mathrm{vis})
					\ r_D(E_\mathrm{vis},\,E_\mathrm{vis}'\,|\,\sigma^2_D),   
\end{equation} 
%............................................................................
via the effective probability
%............................................................................
\begin{equation}
\label{Ansatz}
P_{ee}^\mathrm{eff}(E_\mathrm{vis})\simeq \frac
{\int_{E_T}^\infty dE\ S_T(E_e^\mathrm{mid}+m_e)\ J^{-1}(E)\ P_{ee}(E)\ R_T(E_\mathrm{vis},\, E\,|\,\sigma^2_T)}
{\int_{E_T}^\infty dE\ S_T(E_e^\mathrm{mid}+m_e)\ J^{-1}(E)\  R_T(E_\mathrm{vis},\, E\,|\,\sigma^2_T)}\ .
\end{equation}
%............................................................................
In the limit of no oscillations ($P_{ee}=1=P_{ee}^\mathrm{eff}$), Eq.~(\ref{Mapping2}) reproduces the exact result in Eq.~(\ref{ExactJ}). 
that this recipe can approximately capture the local smearing of $P_{ee}$ implicit in Eq.~(\ref{Exact}) without  
introducing energy biases, as the average recoil effects are accounted for by the mid-recoil approximation. 
Of course, the replacement of $S_\nu(E)$ with the proxy $S_T\left(E_e^\mathrm{mid}(E)+m_e\right)$ introduces an extra 
smearing associated to the latter spectrum, which is absent in the former. This artifact may be expected 
to have marginal effects in the final $S_J$, since the smearing in JUNO acts on an energy scale 
$\sigma_J>\sigma_T$.   Ultimately, the validity of our ansatz relies on numerical tests.

%%%%%%%%%%%%%%%%%%%%%%%%%%%%%%%%%%%%%%%%%%%%%%%%%%%%%%%%%%%%%%%%%%%%%%%%%%%%%%%%%%%%%%%%%%
\begin{figure}[t]
\begin{minipage}[c]{0.9\textwidth}
\includegraphics[width=.48\textwidth]{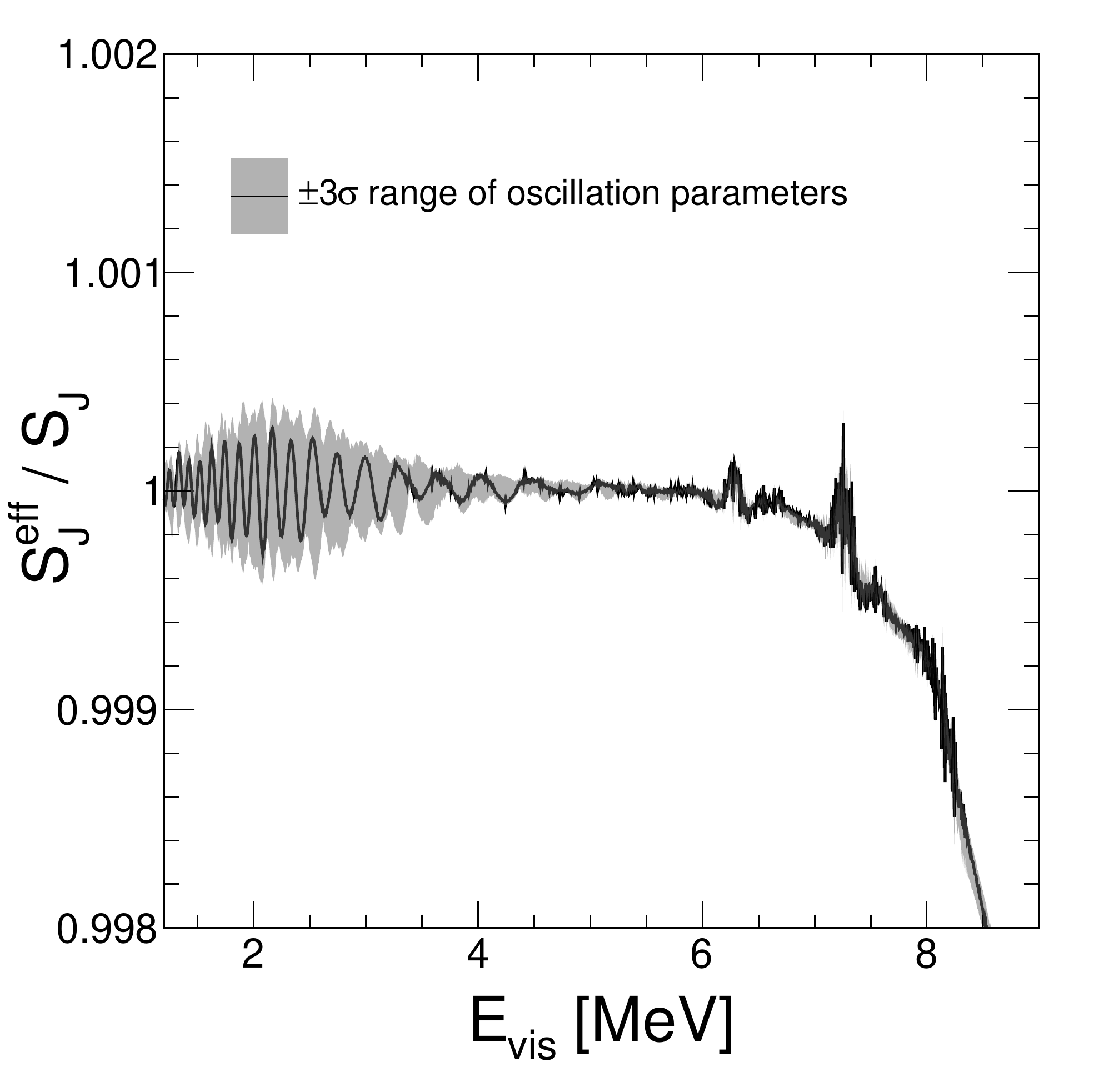}
\caption{\label{Fig_05}
\footnotesize 
Accuracy of the mapping $S_T\to S_J$ in the presence of oscillations:
Ratio of JUNO energy spectra calculated with the ansatz $(S_J^\mathrm{eff})$ and without it ($S_J$).
The solid line is computed for central values of the
oscillation parameters, while the gray band corresponds to the envelope of $\leq 3\sigma$ variations 
in the prior ranges. Normal ordering is assumed. See the text for details.
} 
\end{minipage}
\end{figure}
%%%%%%%%%%%%%%%%%%%%%%%%%%%%%%%%%%%%%%%%%%%%%%%%%%%%%%%%%%%%%%%%%%%%%%%%%%%%%%%%%%%%%%%%%%

Figure~\ref{Fig_05} shows the ratio of the JUNO spectra calculated with the ansatz [$S^\mathrm{eff}_J$ from
Eqs.~(\ref{Mapping2}) and (\ref{Ansatz})] and without the ansatz [$S_J$ from Eq.~(\ref{ExactJ})]. The
underlying neutrino spectrum $S_\nu$ is taken as the reference Oklo spectrum in Fig.~\ref{Fig_01} (left panel). 
The solid line and gray band refer, respectively, to central values of the oscillation parameters and to their $\pm 3\sigma$
variations (applied to both $S^\mathrm{eff}_J$ and $S_J$ at the same time). The ansatz provides numerically accurate
results at the level of few~$\times 10^{-4}$, except in the high-energy tail where it reaches a permill level,
that is anyway insignificant as compared with other sources of
uncertainties (both statistical and systematic) in JUNO, as also discussed below.
Finally, we have tested that the same excellent accuracy in Fig.~\ref{Fig_05} is reached by replacing the
reference spectrum with variant spectra, 
as obtained from the Oklo toolkit by changing the nuclear inputs 
within their uncertainties (not shown).

%%%%%%%%%%%%%%%%%%%%%%%%%%%%%%%%%%%%%%%%%%%%%%%%%%%%%%%%%%%%%%%%%%%%%%%%%%%%%%%%%%%%%%%%%%
\begin{figure}[b]
\begin{minipage}[c]{0.9\textwidth}
\includegraphics[width=.48\textwidth]{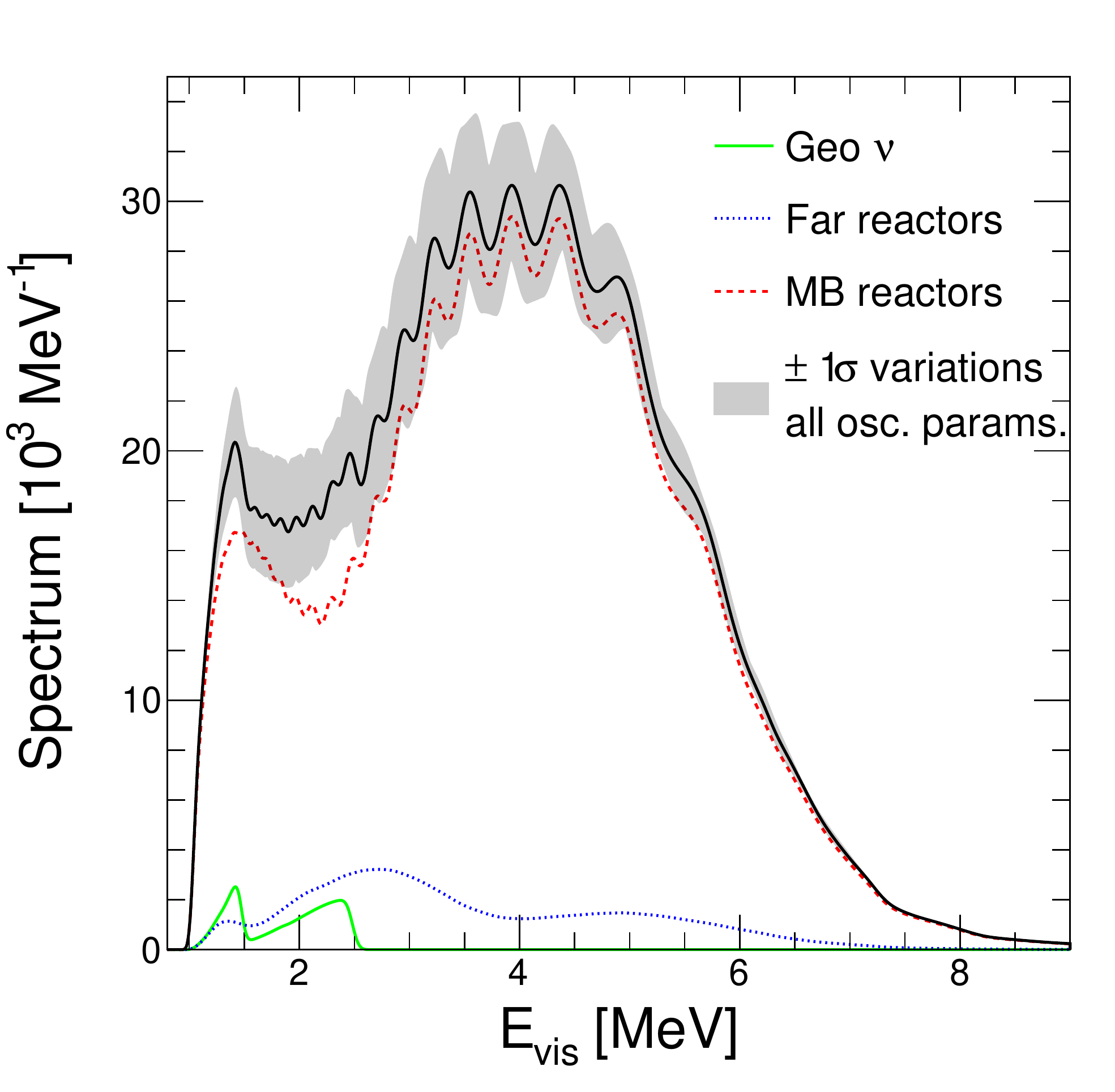}
\caption{\label{Fig_06}
\footnotesize 
Absolute energy spectrum of IBD events expected after 5 years in JUNO,  for oscillation
parameters taken at their central value (black solid line) or left free within $\leq 1\sigma$  
(gray envelope). Normal ordering is assumed. 
The breakdown of the total spectrum in its three components  (MBL reactors, far reactors, and geoneutrinos) 
is also shown. The red dashed line corresponds to the spectrum $S_J$ discussed in the text.} 
\end{minipage}
\end{figure}
%%%%%%%%%%%%%%%%%%%%%%%%%%%%%%%%%%%%%%%%%%%%%%%%%%%%%%%%%%%%%%%%%%%%%%%%%%%%%%%%%%%%%%%%%%

%%%%%%%%%%%%%%%%%%%%%%%%%%%%%%
\section{Neutrino oscillometry in JUNO: single spectrum}
\label{Sec:Single}
%%%%%%%%%%%%%%%%%%%%%%%%%%%%%%%%

We present and discuss a prospective analysis of JUNO in terms of sensitivity to mass ordering and 
of precision determination of oscillation parameters, building upon our previous work \cite{Capozzi:2015bpa}.
Here we use a single input spectrum, namely, the reference Oklo one as shown 
in the previous Sections.
Bundles of variant spectra and their effects will be considered in the next Section.
The main purpose of this updated analysis is to further test the previous ansatz and to discuss the impact
of changes in the reference oscillation parameters and other systematics. 
TAO does not play a specific role herein, except for
providing a reference spectrum $S_T$ for the $S_T\to S_J$ mapping, when the ansatz is used.

\subsection{Ingredients of the analysis}

Figure~\ref{Fig_06} shows the observable JUNO spectrum $S_J$ expected in the presence of oscillations from the 
Taishan and Yangjiang reactor sources    
(dashed red line) plus the background components expected from farther reactors (blue dotted line)
and U+Th geoneutrinos (green solid line).% 
%--------------------------
\footnote{The double-peaked (U+Th) structure of the geo-$\nu$ spectrum is a peculiar realization 
of sawtooth substructures in summation spectra.}    
%---------------------------
The total spectrum (black solid line) is endowed with 
a gray band, representing the envelope of variations of the oscillation parameters
within their prior $1\sigma$ ranges. 
All curves refer to 5 years of data taking ($\sim 10^5$ JUNO events), assuming the same
normalization factors for the various components
as discussed in  \cite{Capozzi:2015bpa}, to which we refer the reader for details not repeated herein.

We focus here on the inputs that differ from \cite{Capozzi:2015bpa}. The central values (and to some extent 
the errors) of the oscillation parameters in Eqs.~(\ref{prior1})--(\ref{prior4}) have changed, in particular
for the mass splittings (about $+1\sigma$ for $\Delta m^2_{ee}$ and $-1\sigma$ for $\delta m^2_{ee}$).
Concerning $\Phi$, we use the reference neutrino flux from the Oklo toolkit, corresponding to the
neutrino spectrum $S_\nu=\Phi\sigma_\nu$ in Fig.~\ref{Fig_01} (left panel). Note that, in this Section, we do not 
attach uncertainties to the fine structures of $\Phi$, that will be separately addressed in 
Sec.~\ref{Sec:Ensemble}. However, we do include large-scale (smooth) uncertainties of the flux shape in the form
of polynomial deviations $\Phi'/\Phi$, as well as energy-scale systematics in the form of polynomial
deviations $E'/E$, adopting the same methodology as in \cite{Capozzi:2015bpa} but with narrower error bands.

%%%%%%%%%%%%%%%%%%%%%%%%%%%%%%%%%%%%%%%%%%%%%%%%%%%%%%%%%%%%%%%%%%%%%%%%%%%%%%%%%%%%%%%%%%
\begin{figure}[t]
\begin{minipage}[c]{0.9\textwidth}
\includegraphics[width=.7\textwidth]{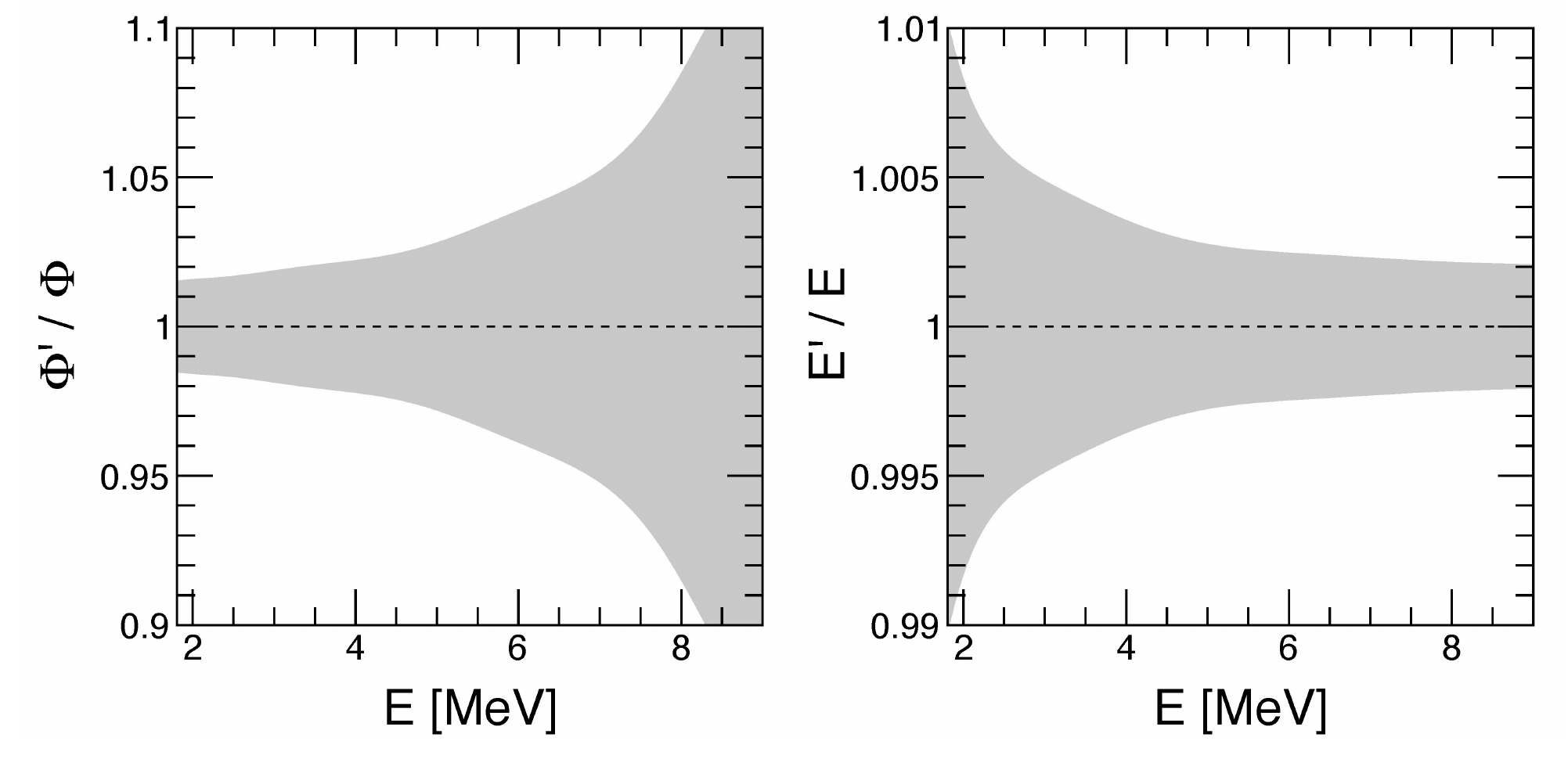}
\caption{\label{Fig_07}
\footnotesize 
Error bands ($\pm 1\sigma$) assumed for flux-shape variations $\Phi'/\Phi$ (left panel) and energy-scale variations
$E'/E$ in JUNO. } 
\end{minipage}
\end{figure}
%%%%%%%%%%%%%%%%%%%%%%%%%%%%%%%%%%%%%%%%%%%%%%%%%%%%%%%%%%%%%%%%%%%%%%%%%%%%%%%%%%%%%%%%%%

Figure~\ref{Fig_07} shows our default $\pm 1\sigma$ error bands  for $\Phi'/\Phi$ and $E'/E$ variations (left and right panels,
respectively). Here we reduce the width of the $\Phi'/\Phi$ band to $2/3$ of the previously adopted one in \cite{Capozzi:2015bpa},
because: (1) at low energy, the normalization error (that sets the lower limit to the width) has been 
reduced from $\sim2.3\%$  \cite{Capozzi:2015bpa} to $\sim1.5\%$  \cite{Adey:2018qct};
at high energies, prospective analyses of the flux shape reconstruction in TAO \cite{Cao-2019} give reasons 
for moderate optimism. The $E'/E$ error band is taken from \cite{Adey:2019zfo} (see Fig.~18 therein), with an 
appreciable reduction (roughly by a factor $1/2$) with respect to \cite{Capozzi:2015bpa}.

%%%%%%%%%%%%%%%%%%%%%%%%%%%%%%%%%%%%%%%%%%%%%%%%%%%%%%%%%%%%%%%%%%%%%%%%%%%%%%%%%%%%%%%%%%
\begin{figure}[t]
\begin{minipage}[c]{0.9\textwidth}
\includegraphics[width=.42\textwidth]{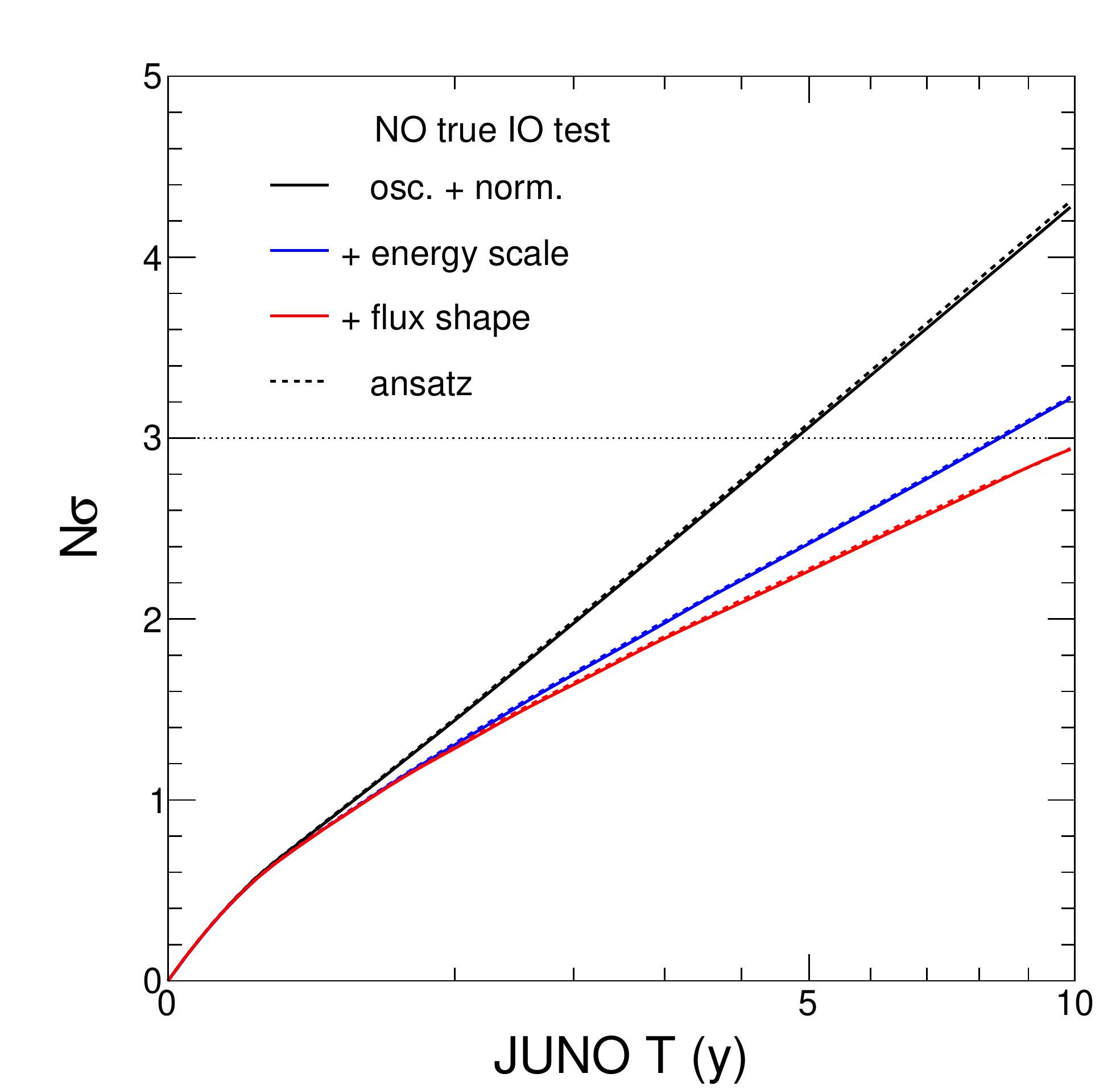}
\caption{\label{Fig_08}
\footnotesize 
JUNO analysis: Statistical significance of the rejection of inverted ordering (IO, test hypothesis)
with respect to normal ordering (NO, true hypothesis), as a function of the live time $T$,
including different sets of systematics: oscillation and normalization uncertainties (black),
plus energy-scale uncertainties (blue) plus flux-shape uncertainties (red). Dashed lines refer to the calculation of
the JUNO spectrum by mapping the TAO spectrum (ansatz discussed in the text).} 
\end{minipage}
\end{figure}
%%%%%%%%%%%%%%%%%%%%%%%%%%%%%%%%%%%%%%%%%%%%%%%%%%%%%%%%%%%%%%%%%%%%%%%%%%%%%%%%%%%%%%%%%%

\subsection{Sensitivity to mass ordering}

Following~\cite{Capozzi:2015bpa}, we perform a least-squares analysis of the JUNO sensitivity to mass ordering, up to 10 years
of data taking. We remind that our complete $\chi^2$ function
for JUNO  is defined as
%-----------
\begin{equation}
\chi^2_\mathrm{JUNO} = \chi^2_{\mathrm{stat}}+\chi^2_{\mathrm{par}}+\chi^2_{\mathrm{sys}},
\end{equation}
%----------
where: the first term includes statistical errors only; the second term includes penalties for variations of the oscillation
parameters, governed by the priors in Rqs.~(\ref{prior1})--(\ref{prior4}); the third term contains normalization errors for 
the geo-$\nu$ Th and U fluxes, normalization and (polynomial) shape systematics for the reactor fluxes, and
(polynomial) energy-scale systematics. The second and third term contain up to $N_\mathrm{sys}=18$ systematics,
treated as nuisance parameters that are marginalized away in the $\chi^2_J$ minimization \cite{Capozzi:2015bpa}.
The analysis is performed by progressively including such nuisance parameters: 
(1) oscillation parameters and normalizations (osc.~+~norm.), $N_\mathrm{sys}=7$; (2) plus energy scale variations,
$N_\mathrm{sys}=13$;
(3) plus flux shape variations, $N_\mathrm{sys}=18$. Normal (inverted) ordering is assumed as true (test) hypothesis.

Figure~\ref{Fig_08} shows the results of the JUNO analysis in terms of standard deviations 
[$N_\sigma = \sqrt{\Delta \chi^2 (\mathrm{IO-NO})}$] as a function of the detector live time $T$, with tic marks scaling as $\sqrt{T}$.
Solid lines refers to the standard calculation of $S_J$ from $S_\nu$, while dashed lines to the approximate
$S_T\to S_J$ mapping; the excellent agreement corroborates the validity of our ansatz.  
The statistical rejection of the wrong IO 
reaches $2$--$3\sigma$ in $5$--$10$~years, depending on systematic errors. 
Note that systematics do not seem to saturate the sensitivity to mass ordering even with 10-years data.
Also note that this sensitivity is reduced more by energy-scale uncertainties 
than by flux-shape ones. Therefore, it will be important to ensure that the
energy calibration in JUNO can achieve the same (or better) level of accuracy reached in \cite{Adey:2018qct}.

It is useful to compare this Fig.~\ref{Fig_08} with the analogous Fig.~7 in \cite{Capozzi:2015bpa}.
It turns out that the curve $N_\sigma(T)$ including the full set of systematics (red curve) is almost unaltered, despite
the previously discussed reduction in energy scale and flux shape systematics. The (surprising to us)
explanation is that the benefits of this error reduction happen to be accidentally compensated by ``unlucky''  changes in
the central values of the oscillation parameters. Further understanding can be gained by focusing on the
case with ``osc.~+~norm.'' errors only (black curve  in Fig.~\ref{Fig_08}) for a fixed live time $T=5$~years.

%%%%%%%%%%%%%%%%%%%%%%%%%%%%%%%%%%%%%%%%%%%%%%%%%%%%%%%%%%%%%%%%%%%%%%%%%%%%%%%%%%%%%%%%%%
\begin{figure}[b]
\begin{minipage}[c]{0.9\textwidth}
\includegraphics[width=.76\textwidth]{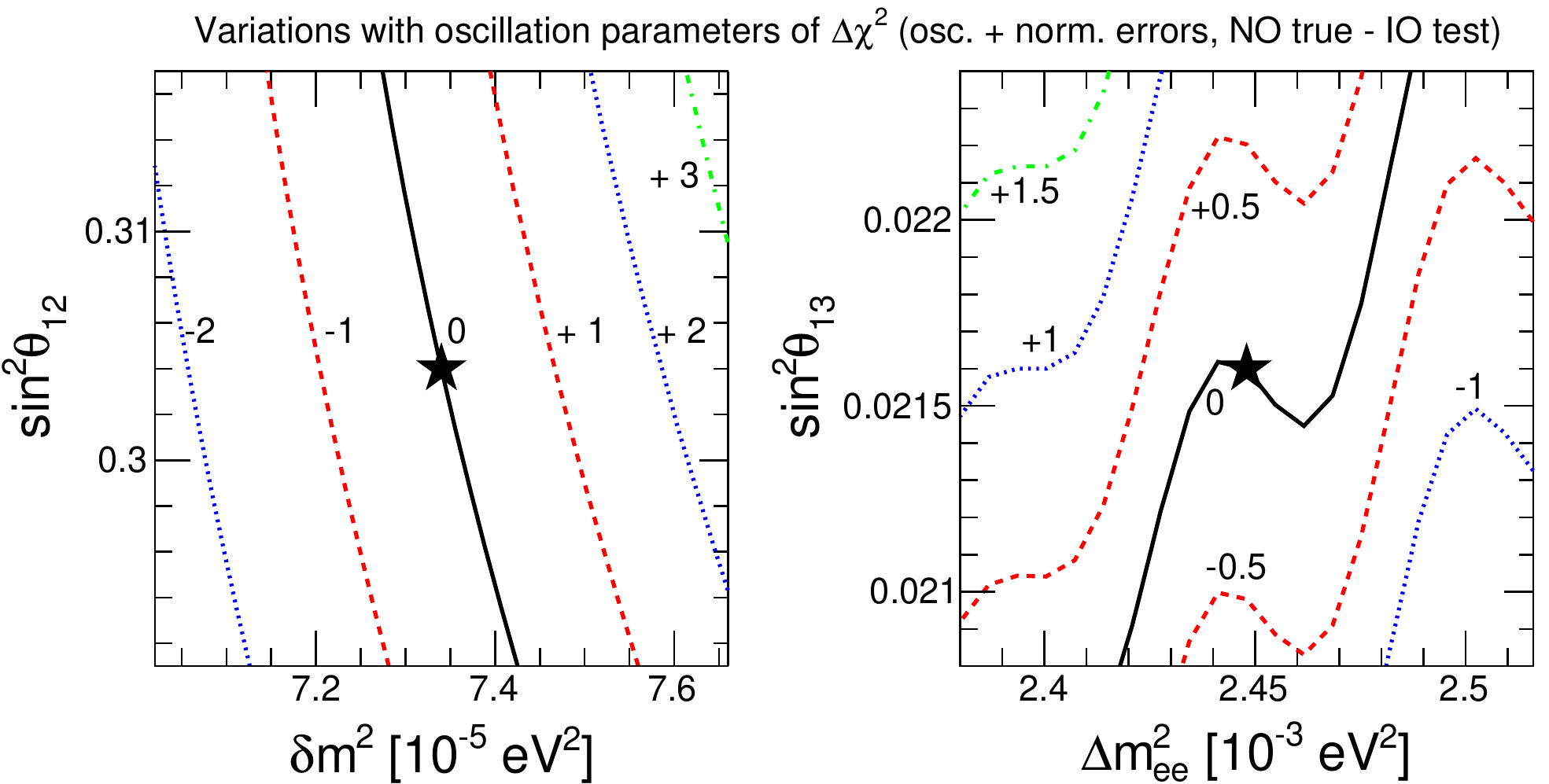}
\caption{\label{Fig_09}
\footnotesize 
JUNO analysis: Isolines of $\Delta\chi^2$ variations for the test of IO (assuming true NO), including only
oscillation and normalization errors. The left and right planes are charted by $(\delta m^2,\,\sin^2\theta_{12})$
and $(\Delta m^2_{ee},\,\sin^2\theta_{13})$, respectively. The
central values of the oscillation parameters are marked by a star. Results refer to $T=5$~years.} 
\end{minipage}
\end{figure}
%%%%%%%%%%%%%%%%%%%%%%%%%%%%%%%%%%%%%%%%%%%%%%%%%%%%%%%%%%%%%%%%%%%%%%%%%%%%%%%%%%%%%%%%%%

Figure~\ref{Fig_09} shows how the $\Delta\chi^2(\mathrm{IO-NO})$ changes by varying
the central values of the oscillation parameters, with respect to those reported in 
Eqs.~(\ref{prior1})--(\ref{prior4}) and marked by a star. The left panel
shows $\Delta\chi^2$ variations (isolines) in the plane $(\delta m^2,\,\sin^2\theta_{12})$ for fixed
$(\Delta m^2_{ee},\,\sin^2\theta_{13})$, and viceversa in the right panel. The coordinates span
the $\pm2\sigma$ ranges for the mass splitting and  $\pm1\sigma$ ranges for the mixing angles,
in the units of Eqs.~(\ref{prior1})--(\ref{prior4}). The $\Delta\chi^2$ value increases
noticeably by increasing $\delta m^2$ or by decreasing $\Delta m^2$; in other words, 
the mass ordering test in JUNO improves when the ratio $\rho = \delta m^2/\Delta m^2_{ee}$ increases, even if by small
amounts (conversely, the mass ordering would become eventually untestable for vanishing $\rho$).
Note that a (more modest) increase of $\Delta \chi^2$ is also obtained by increasing either $\sin^2\theta_{12}$ or 
$\sin^2\theta_{13}$ and thus the oscillation amplitude(s), as it can be generally expected in oscillation
searches.

It turns out that, with respect to  \cite{Capozzi:2015bpa}, the central values of all four oscillation parameters in 
Eqs.~(\ref{prior1})--(\ref{prior4}) have accidentally changed in ``unlucky'' directions,  
lowering $\Delta \chi^2$ by  about $3.5$ units  for the case of ``osc.~+~norm.'' uncertainties.
As anticipated, for the analysis including all the uncertainties,
this drop is almost exactly compensated (once more, accidentally) by the reduction
of energy-scale and flux-shape systematics. Similar results have been obtained in the case where the true ordering
is inverted and NO is tested (not shown).
In conclusion, the JUNO rejection of the wrong mass ordering depends, in a nonnegligible  way, on
the central values of the oscillation parameters.

%%%%%%%%%%%%%%%%%%%%%%%%%%%%%%%%%%%%%%%%%%%%%%%%%%%%%%%%%%%%%%%%%%%%%%%%%%%%%%%%%%%%%%%%%%
\begin{figure}[b]
\begin{minipage}[c]{0.9\textwidth}
\includegraphics[width=.40\textwidth]{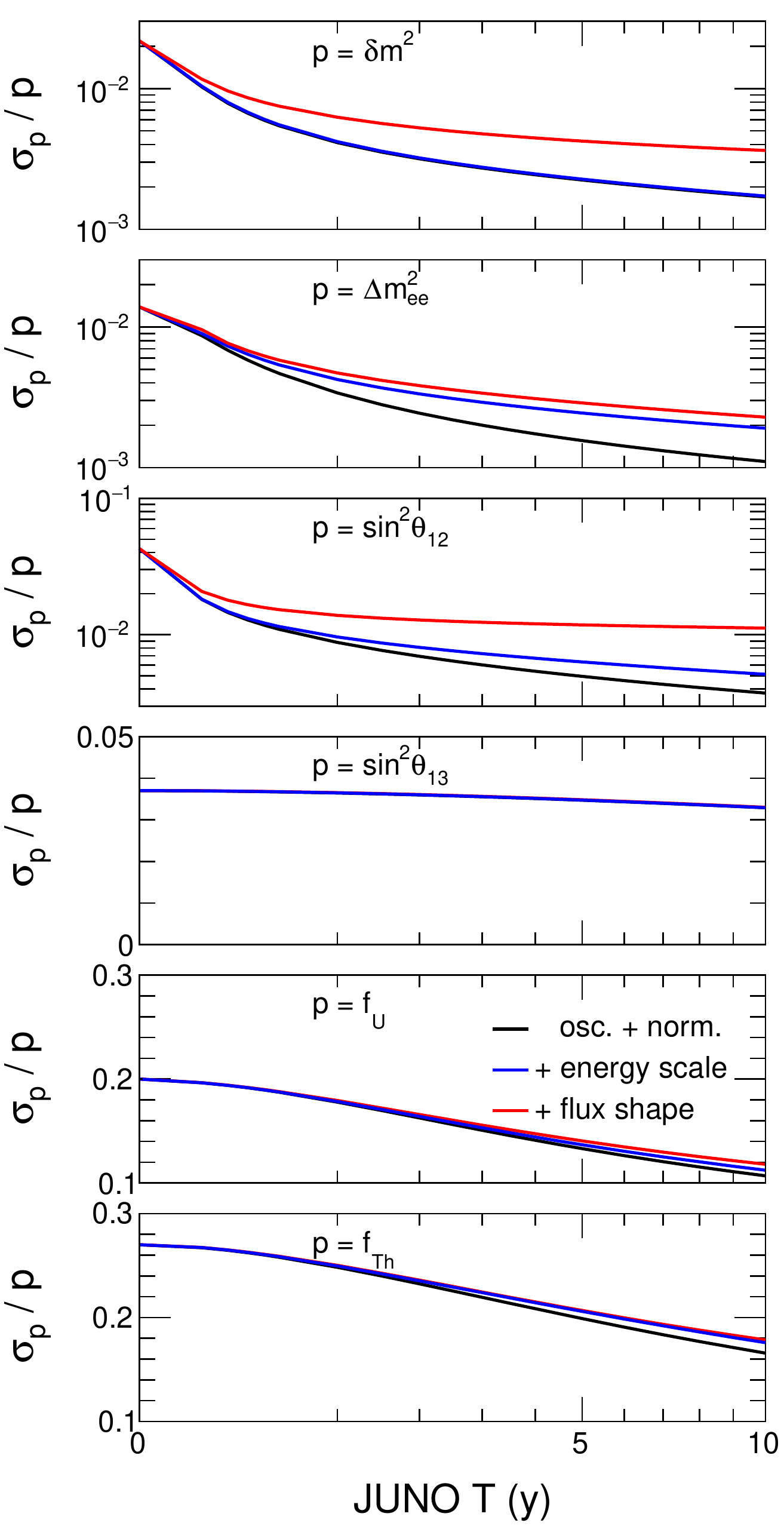}
\caption{\label{Fig_10}
\footnotesize 
JUNO analysis: fractional accuracy $\sigma_p/p$ as a function of live time $T$ for six 
measurable parameters $p$. The line color code is the same as in Fig.~\ref{Fig_08}, and the abscissa
also scales as $\sqrt{T}$. From top
to bottom, the results refer to the squared mass splittings $\delta m^2$ and $\Delta m^2_{ee}$, 
the mixing angles $\sin^2\theta_{12}$ and $\sin^2\theta_{13}$, and the geoneutrino flux
normalization factors $f_\mathrm{U}$ and $f_\mathrm{Th}$. The same results are obtained by using
the $S_T\to S_J$ mapping ansatz (not shown).
} 
\end{minipage}
\end{figure}
%%%%%%%%%%%%%%%%%%%%%%%%%%%%%%%%%%%%%%%%%%%%%%%%%%%%%%%%%%%%%%%%%%%%%%%%%%%%%%%%%%%%%%%%%%

\subsection{Accuracy of oscillation parameters}

Eventually, at least three oscillation parameters ($\delta m^2,\,\Delta m^2_{ee},\,\theta_{12}$) will be 
very precisely measured by JUNO itself. Figure~\ref{Fig_10} shows the time evolution (in JUNO) of the fractional accuracy 
$\sigma_p/p$ for each of six parameters $p$, namely, from top to bottom: the two mass splittings, the two mixing angles, and 
the U and Th geoneutrino flux normalizations ($f_\mathrm{U}$ and $f_\mathrm{Th}$). For each parameter, it is understood that the
others are marginalized away in the analysis.
At $T=0$, the oscillation parameter errors are set by 
 Eqs.~(\ref{prior1})--(\ref{prior4}), while for the geo-$\nu$ fluxes we assume the priors in  \cite{Capozzi:2015bpa},
 $f_\mathrm{U}=1\pm 0.20$ and $f_\mathrm{Th}=1\pm 0.27$. The color sequence for the curves (red, blue and black for growing sets of systematics)
 is the same as in Fig.~\ref{Fig_08}.
After a live time of 5 years, the accuracy of ($\delta m^2,\,\Delta m^2_{ee},\,\theta_{12}$) will improve by
factors of about $(6,\,6,\,4)$, respectively---or better, if some systematics che be further reduced. 
For the pair ($\delta m^2,\,\Delta m^2_{ee})$, that
governs the ``slow'' oscillations in the JUNO spectrum, flux-shape 
uncertainties are more important than energy-scale ones, and viceversa for $\Delta m^2_{ee}$
that governs the ``fast'' oscillations.  A moderate reduction
of the prior errors will be obtained for geoneutrino fluxes, with little dependence on systematics. 
Concerning $\theta_{13}$, the current experimental error
will only be marginally improved. Finally, we have repeated the analysis by
using the $S_T\to S_J$ mapping ansatz, obtaining the same results with insignificant deviations (not shown).

%%%%%%%%%%%%%%%%%%%%%%%%%%%%%%%%%%%%%%%%%%%%%%%
\section{Neutrino oscillometry in JUNO: ensembles of spectra}
\label{Sec:Ensemble}
%%%%%%%%%%%%%%%%%%%%%%%%%%%%%%%%%%%%%%%%%%%%%%%

Summation calculations of reactor spectra have come a long way since the pioneering works 
\cite{King:1958zz,Avignone,Davis:1979gg,Avignone:1980qg}. Modern realizations are based
on thousands of nuclear input data on decay yields $Y_i$, endpoints $Q_j$ and branching ratios $b_k$, 
together with their uncertainties and possible covariances  \cite{Sonzogni:2017voo,Sonzogni-IAEA}.
However, as mentioned in the Introduction, even the most refined summation spectra do not match well 
current reactor data, suggesting that some nuclear (experimental or theoretical) ingredients may be missing.
Significant work is still needed to reach consensus on satisfactory spectra with realistic uncertainties 
and correlations \cite{IAEA-Vienna,Sonzogni-AAP}, with TAO  providing important  benchmarks in the future \cite{TAO-CDR}.

With all these caveats, we perform an exploratory analysis of the effect of ``known'' 
nuclear input uncertainties on the spectral substructures  through the Oklo toolkit \cite{Oklo}. We remind that the Oklo
 code contains 
$(4306,\, 6609,\, 6804)$ values for $(Y_i,\,Q_j,\,b_k)$, respectively, for a total of $N_d=17,719$ input data,
together with their quoted uncertainties (taken as uncorrelated). These huge numbers prevent usual $\chi^2$ analyses
of variant spectra, in terms of marginalization over nuisance parameters.  
Alternatively, we generate 
ensembles of $N$ neutrino spectra $\{S_\nu^n(E)\}_{n=1,\dots,N}$, by randomly varying all or some nuclear inputs within their
uncertainties. We also compute 
the associated TAO spectra $\{S_T^n(E_\mathrm{vis})\}$, that are then  mapped to obtain JUNO spectra  $\{S_J^n(E_\mathrm{vis})\}$
(where we drop the superscript ``eff'' for simplicity).

We test how these variants affect the JUNO oscillation analysis, and how well they can be distinguished by TAO,
by scanning appropriate $\chi^2$ functions over the whole spectral set(s).  
We recover, through an independent $\chi^2$
analysis, the results obtained in \cite{Danielson:2018tzi} 
through a Fourier analysis, namely, that ``known'' substructure uncertainties do not appear to
pose a threat to precision oscillometry in JUNO. However, the quantification of this result
is not trivial, and some subtle problems in the statistical analysis will  be highlighted.
We shall also comment on the issue of possible ``unknown'' small-scale uncertainties, as raised e.g.\ in \cite{Forero:2017vrg}
and in \cite{Cheng:2020ivh,Ciuffoli:2019nli}.

\subsection{Changing all nuclear input uncertainties: spectrum metric and (under)sampling issues}

In our first exercise with spectral variants,
we have generated an ensemble of $N=10^5$ neutrino spectra $S_\nu^n$ (and 
associated TAO spectra $S_T^n$) by $N$ extractions of random values $s^n_i$ for all the $i=1,\dots,N_d$  inputs
at the same time, assuming uncorrelated gaussian distributions for the quoted uncertainties $\sigma_i$.
 At each extraction, branching ratios for each decay are renormalized
by an overall factor to ensure unitarity ($\sum_k b_k=1$).
All variant spectra $S_T^n$ are normalized to the same area as the
reference spectrum $S_T$, in order to emphasize shape variations. 

\begin{figure}[t]
\begin{minipage}[c]{0.9\textwidth}
\includegraphics[width=.49\textwidth]{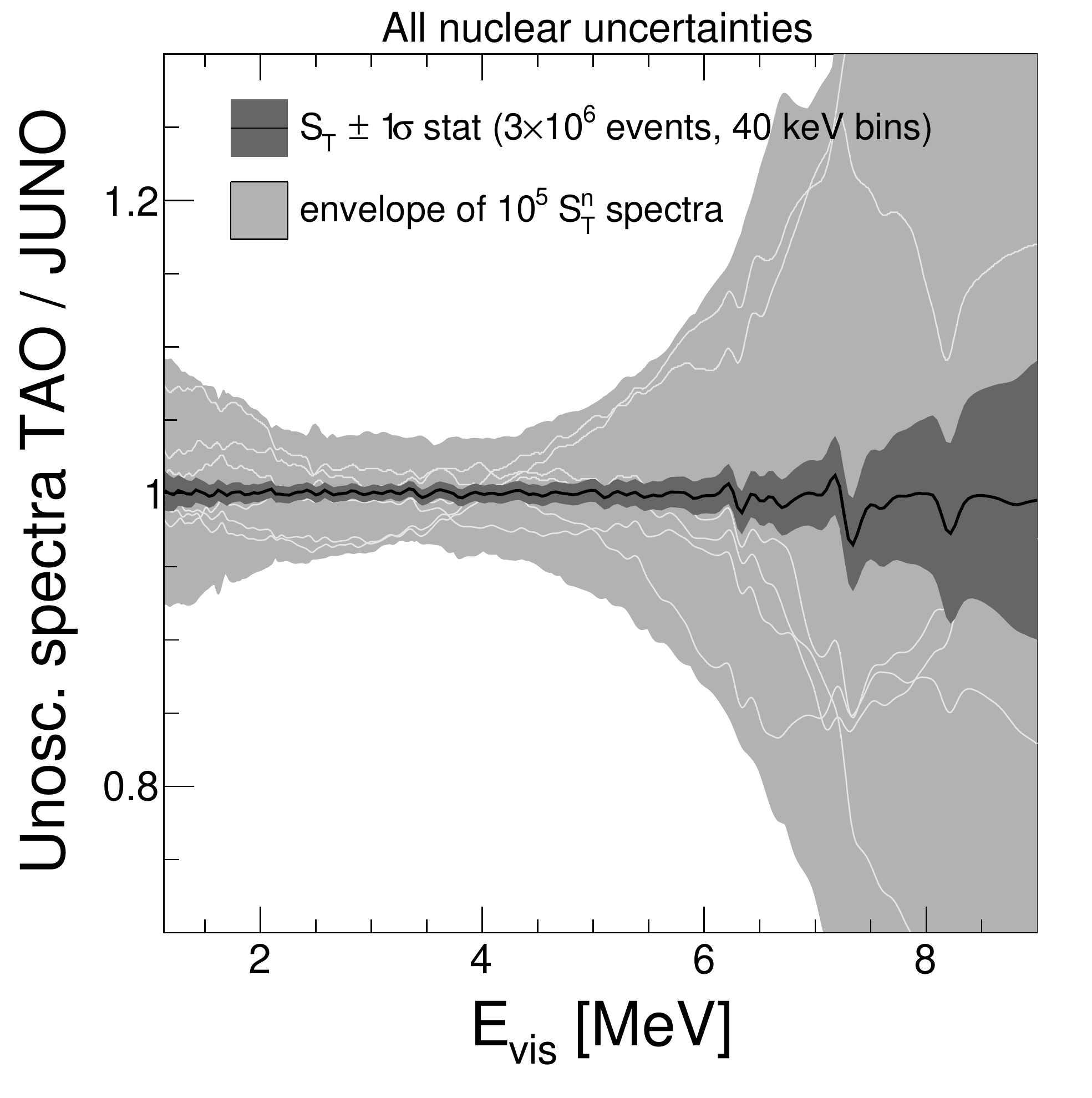}
\caption{\label{Fig_11}
\footnotesize Spectral ensembles in TAO. Solid line with dark gray band: TAO reference spectrum $S_T$ with its statistical errors,
assuming $3\times 10^6$ IBD events and 40 keV bins. 
Light gray  band: Envelope of spectra $\{S^n_T\}_{n=1,\dots,N}$ at TAO, as obtained by $N=10^5$ extractions of gaussian-distributed 
values for all the $N_d=17,719$ nuclear input uncertainties in the Oklo toolkit, and normalized to same area as $S_T$.
A few individual variants are also shown (very light gray curves). 
All spectra are divided by the unoscillated reference JUNO spectrum $S_J$, in order to show fine structures.
} 
\end{minipage}
\end{figure}

Figure~\ref{Fig_11} shows $S_T$ (solid line) with its statistical errors (dark gray band), 
assuming $3\times 10^{6}$ IBD events in TAO, and $40$ keV bins. Also shown is the envelope of all the $S_T^n$ variant
spectra (light gray band), and a few individual variants (very light gray curves).  
All spectra are divided by the unoscillated JUNO spectrum $S_J$, analogously to Fig.~\ref{Fig_03}. 
Since the light gray band is rather large, one may expect that at least some 
spectral variants within the envelope can play a role in the TAO and JUNO data analyses.
The surprising outcome is that only the reference $S_T$ matters in our exercise, for subtle reasons
that we could not anticipate.

%%%%%%%%%%%%%%%%%%%%%%%%%%%%%%%%%%%%%%%%%%%%%%%%%%%%%%%%%%%%%%%%%%%%%%%%%%%%%%%%%%%%%%%%%%
%%%%%%%%%%%%%%%%%%%%%%%%%%%%%%%%%%%%%%%%%%%%%%%%%%%%%%%%%%%%%%%%%%%%%%%%%%%%%%%%%%%%%%%%%%

%%%%%%%%%%%%%%%%%%%%%%%%%%%%%%%%%%%%%%%%%%%%%%%%%%%%%%%%%%%%%%%%%%%%%%%%%%%%%%%%%%%%%%%%%%
\begin{figure}[b]
\begin{minipage}[c]{0.9\textwidth}
\includegraphics[width=.49\textwidth]{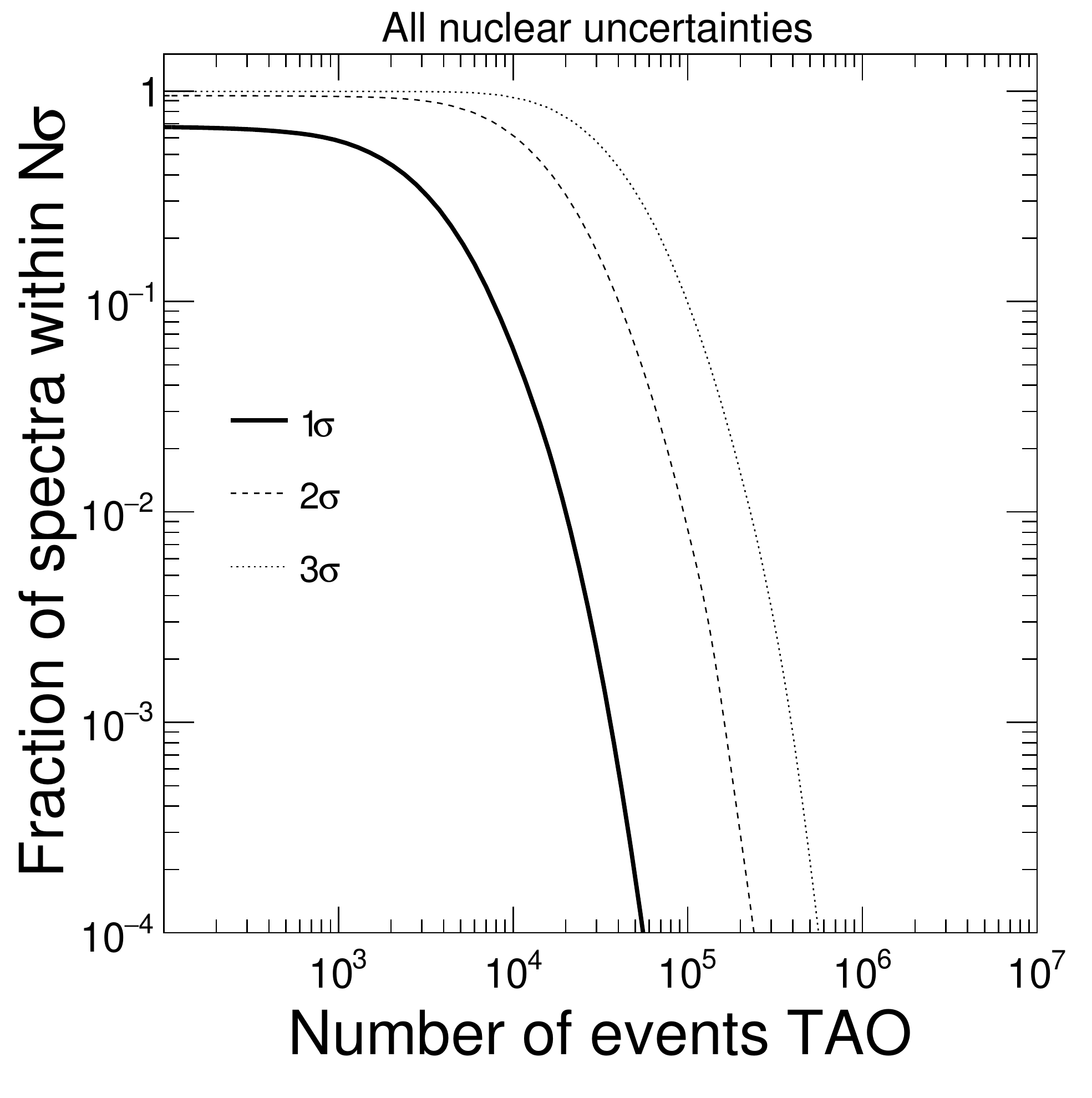}
\caption{\label{Fig_12}
\footnotesize Spectral ensembles in TAO. Fraction of variant spectra $S^n_T$ (generated by changing all nuclear uncertainties)
that survive at $N_\sigma$ when compared to the reference $S_T$ spectrum, as a function of 
accumulated TAO events.  } 
\end{minipage}
\end{figure}
%%%%%%%%%%%%%%%%%%%%%%%%%%%%%%%%%%%%%%%%%%%%%%%%%%%%%%%%%%%%%%%%%%%%%%%%%%%%%%%%%%%%%%%%%%

A first issue is how to define a $\chi^2_S$ metric within the $\{S_T^n\}$ envelope, so that 68\% (95\%) of the spectra fall within
a properly defined $1\sigma$ ($2\sigma$) band etc.\ (with $N_\sigma=\sqrt{\chi^2_S}$) around the reference spectrum $S_T$. 
Note that 
each spectrum $S_T^n$ is endowed with a $\chi^2_n$ value
%...................................
\begin{equation}
\chi^2_n = \sum_{i=1}^{N_d} \; \left(\frac{s^n_i}{\sigma_i}\right)^2
\end{equation}
%..................................
that measures its statistical distance from $S_T$ (having $\chi^2=0$ by definition) in terms of nuclear
input uncertainties. For $N_d\gg 1$, the distribution of $\chi^2_n$ values (not shown)
can be approximated by a gaussian centered at $N_d$ and with variance $2N_d$ \cite{Cowan-PDG}, effectively
starting at $0$ (corresponding to $S_T$) rather than $-\infty$.  Since $S_T$ sits in the tail rather than
at the peak, this distribution does not directly provide a good metric. However one
can construct a proper metric $\chi^2_S$ by integrating
this $\chi^2_n$ distribution from zero up to the fractional area 
corresponding to the desired $N_\sigma$ level. As a result (proof omitted), each spectrum $S_T^n$ is endowed 
with a new $\chi^2_{S,n}$ value given by:
%..................................
\begin{equation}
\label{Penalty}
\chi^2_{S,n}=2\left(
\mathrm{erf}^{-1}\left(
\frac{1}{2}+\frac{1}{2}
\mathrm{erf}
\left(
\frac{\chi^2_n-N_d}{2\sqrt{N_d}}
\right)
\right)
\right)^2
\end{equation}
%..................................
where $\mathrm{erf}^{-1}$ is the inverse error function, and $\chi^2_S=0$ is recovered for $S_T$ in the limit $N_d\gg 1$.

It turns out that if the bands corresponding, e.g., to $\chi^2_{S,n}\leq 1$,~2,~and 3 were plotted in Fig.~\ref{Fig_11},  they would
be insignificantly smaller than the light gray envelope of all the spectra. In other words, by taking spectra with
increasingly high $\chi^2_{S,n}$ (or equivalently $\chi^2_n$), more variant spectral shapes become possible within the
band, while the typical  subtructure amplitudes remain constant and their envelope is not enlarged.  

These
results suggest caution in parametrizing variant spectra as in 
\cite{Forero:2017vrg}, namely, by 
breaking down the envelope in bins and computing uncorrelated standard deviations in each bin, 
for two reasons: (1) the amplitude of deviations does not scale with $N_\sigma$; (2)
by binning, the detailed information about which shapes are (not) allowed by nuclear uncertainties is 
completely lost; in particular, the loss of point-to-point correlations permits more shapes than would be allowed
by nuclear inputs only. In doing so,
``known'' uncertainties are partly replaced by ``unknown'' ones,
allowing substructure amplitudes and shapes beyond those pertaining to compiled nuclear inputs.

A second issue  concerns the fraction of spectra $\{S_T^n\}$ that survives the comparison with prospective TAO data. We 
consider a simplified $\chi^2$ analysis for TAO, where each spectrum $S_T^n$ is compared with the reference one $S_T$  
in terms of statistical errors,
plus one nuisance normalization parameter $\lambda$ ($S_T^n\to \lambda S_T^n$, assuming $\sigma_\lambda/\lambda=1.5\times 10^{-2}$), 
in addition to $\chi^2_{S,n}$ that embeds
nuclear errors: 
%.......................
\begin{equation}
\label{TAOchi2}
\chi^2_{\mathrm{TAO},n} = \chi^2_{\mathrm{stat},n} + \chi^2_{\mathrm{norm},n} + \chi^2_{S,n},
\end{equation}
%.......................
where for $\chi^2_{\mathrm{stat},n}$ we adopt the limit of infinite bins \cite{Ge:2012wj,Capozzi:2013psa,Capozzi:2015bpa}, 
that provides a very good approximation to the binned case. Within the ensemble $\{S_T^n\}$, the fraction of spectra 
allowed at $N_\sigma$ by TAO data (defined by $\chi^2_{\mathrm{TAO},n}\leq N^2_\sigma$) is a function of the
TAO exposure. With $\sim 3\times 10^6$ IBD events expected in TAO after $\sim 5$~years, we unexpectedly find that none of
the $10^5$ synthetic spectra survives, even at $N_\sigma=3$ level: they are all rejected with respect to the reference
spectrum $S_T$. It turns out that the good TAO energy resolution
is sufficient to distinguish spectra $S_T^n$ that differ from $S_T$ by a few substructures, 
even with much less than $\sim 3\times 10^6$ events.

Figure~\ref{Fig_12} shows how well TAO selects spectra in the ensemble $\{S_T^n\}$, as a function of the 
total number of collected IBD events. The three curves represent the fraction of spectra
 that survives at $N_\sigma =1$, 2 and 3. For no TAO events  
these fraction correspond by construction, respectively, to 0.68, 0.95 and 0.997. By increasing
the number of TAO events, these fractions drop rapidly.  When the surviving fractions drop below
$10^{-4}$ (not shown), the curves break down  because 
only a handful of the $10^{5}$ spectra---and ultimately none of them---is allowed; this
happens well below $3\times 10^6$ events in TAO. The results in Fig.~\ref{Fig_12} suggest that, when all the
nuclear input uncertainties ($N_d=17,719$) are randomly varied,  generating $10^5$ synthetic spectra by random 
extractions  
is not enough to densely sample the $\infty^{N_d}$-dimensional set of possible variant spectra: 
orders of magnitude more extractions  would be needed to  obtain a few spectra $S_T^n$ really close
to $S_T$ within statistical uncertainties. 

A third statistical issue, connected with the last one just discussed,
concerns the JUNO sensitivity to mass ordering. We have repeated 
the prospective JUNO data analysis in Sec.~\ref{Sec:Single} by mapping the spectral ensemble  
$\{S_T^n\}\to \{S_J^n\}$ for any set of oscillation parameters. 
In particular, assuming NO and the reference $S_J$ as the true hypothesis, we have tested
the wrong IO not only via $S_J$ but also scanning the $10^5$ spectra $S_J^n$ (with and without
adding the term $\chi^2_{S,n}$).% 
%---------------------------------
\footnote{
Our computing resources are saturated for $O(10^5)$ replicas of prospective JUNO data analyses, hence 
the choice of having no more than $10^5$ synthetic spectra.
}
%-------------------------
 We have found no reduction of the sensitivity to the mass ordering as compared with Fig.~\ref{Fig_08}.%
%-------------------------------
\footnote{A similar test with 6 variant spectra in JUNO (rather than $10^5$) was mentioned in \cite{Zhan-2018}.} 
%---------------------------------
These results qualitatively agree with those in \cite{Danielson:2018tzi} (where a Fourier spectral
analysis found that substructures played a little role) but are unexpectedly stronger:
none of the test spectra induces any sensitivity reduction in JUNO. In addition,
we find that also the precision determination of several parameters $p$ as in Fig.~\ref{Fig_10}
remains unaltered. 
Once more, we surmise that 
the ensemble of $10^5$ spectra is not dense enough 
to sample shape variations very close to the reference one. In order to overcome these issues
we construct and test a denser ensemble below.

\subsection{Changing only some nuclear input uncertainties: Suggestions for possible parametrizations}

%%%%%%%%%%%%%%%%%%%%%%%%%%%%%%%%%%%%%%%%%%%%%%%%%%%%%%%%%%%%%%%%%%%%%%%%%%%%%%%%%%%%%%%%%%
\begin{figure}[t]
\begin{minipage}[c]{0.9\textwidth}
\includegraphics[width=.49\textwidth]{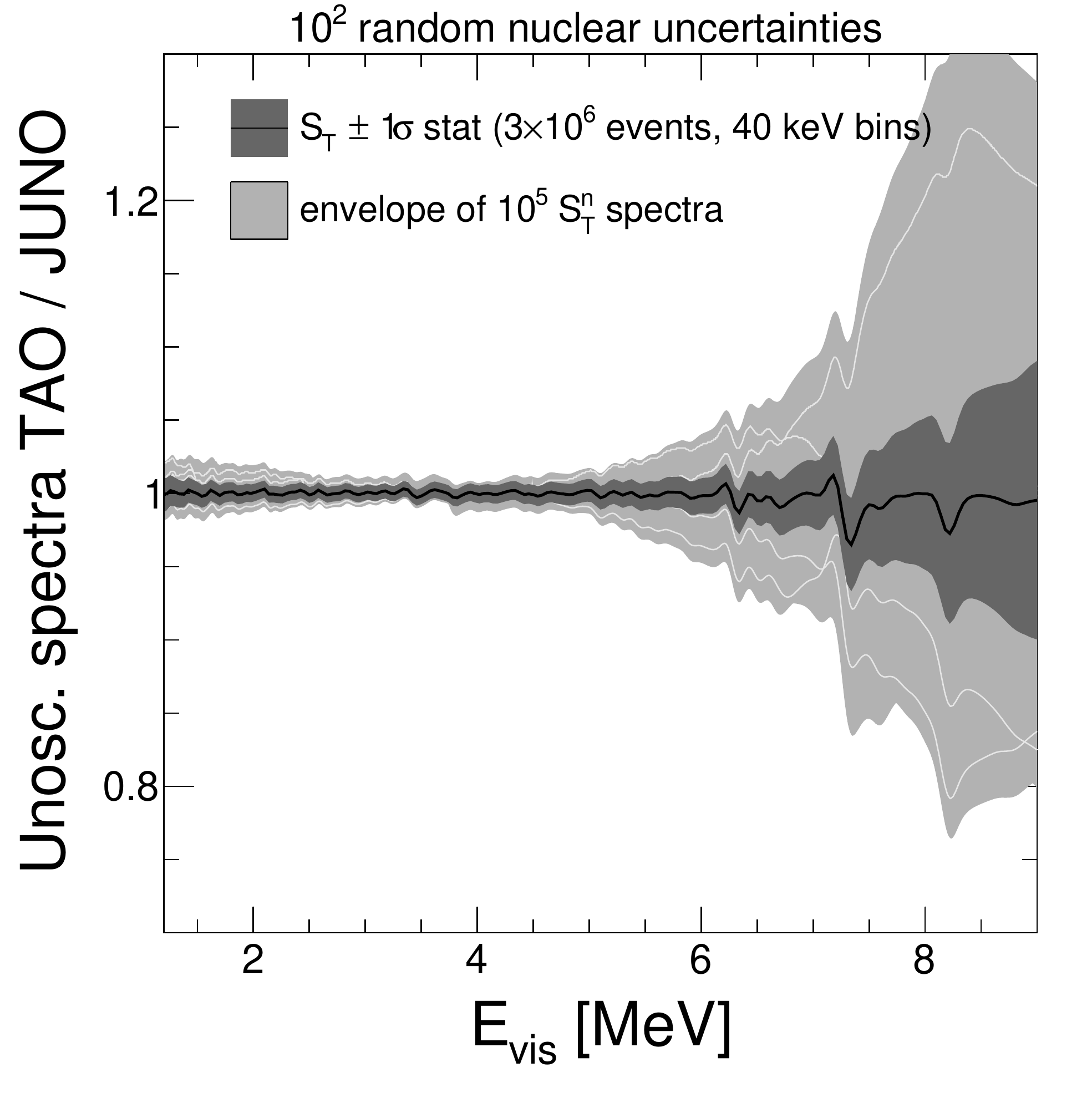}
\caption{\label{Fig_13}
\footnotesize Spectral ensembles in TAO. As in Fig.~\ref{Fig_11}, but with the
light gray  band representing the
envelope of spectra $\{S^n_T\}_{n=1,\dots,N}$ at TAO, as obtained by $N=10^5$ extractions of gaussian-distributed 
values for a random set of
$N'_d=10^2$ (out of $N_d=17,719$) nuclear input uncertainties in the Oklo toolkit.
} 
\end{minipage}
\end{figure}
%%%%%%%%%%%%%%%%%%%%%%%%%%%%%%%%%%%%%%%%%%%%%%%%%%%%%%%%%%%%%%%%%%%%%%%%%%%%%%%%%%%%%%%%%%

%%%%%%%%%%%%%%%%%%%%%%%%%%%%%%%%%%%%%%%%%%%%%%%%%%%%%%%%%%%%%%%%%%%%%%%%%%%%%%%%%%%%%%%%%%
\begin{figure}[b]
\begin{minipage}[c]{0.9\textwidth}
\includegraphics[width=.49\textwidth]{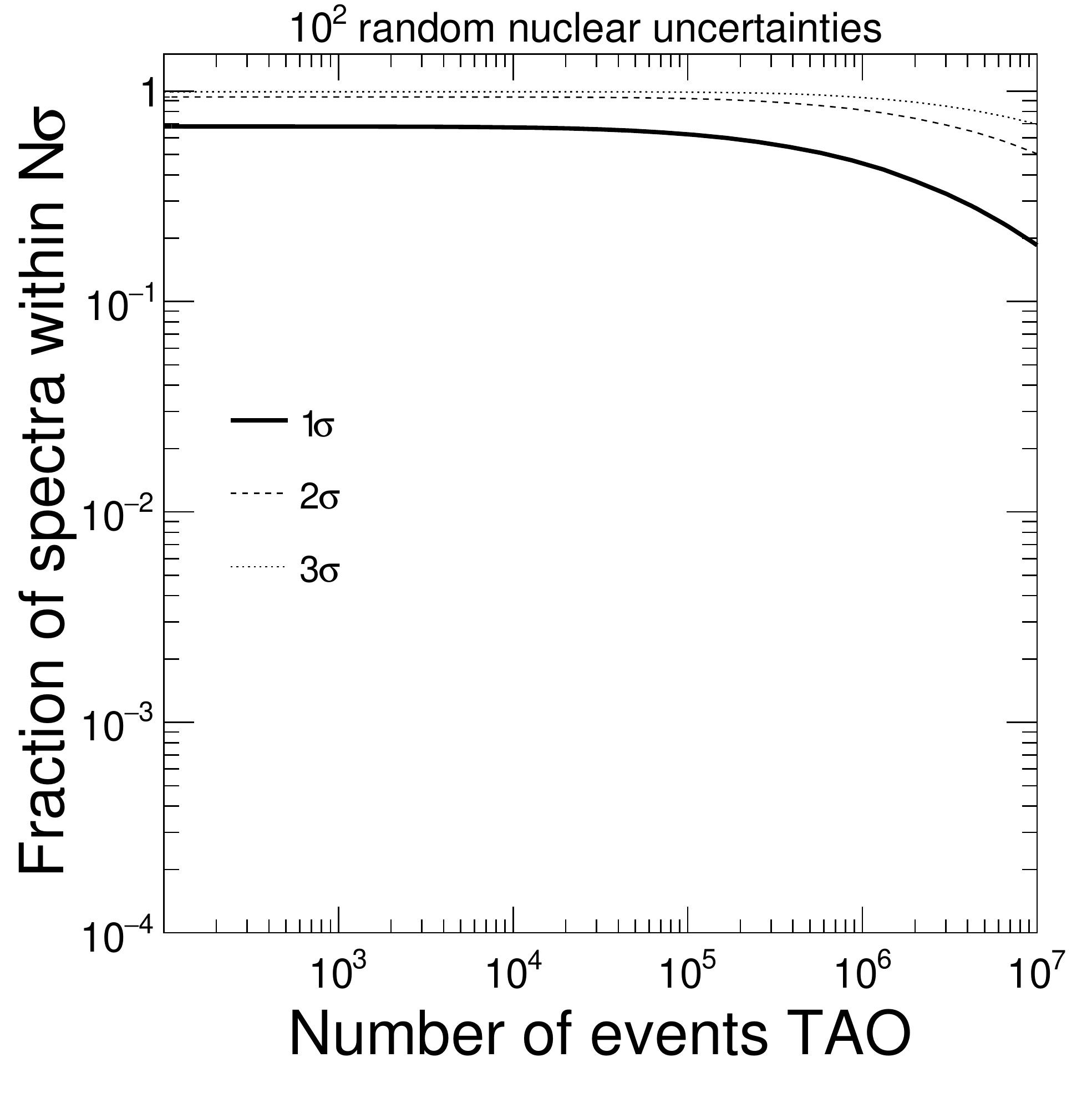}
\caption{\label{Fig_14}
\footnotesize Spectral ensembles in TAO. As in Fig.~\ref{Fig_12}, but considering only
a random subset of 100 nuclear uncertainties. } 
\end{minipage}
\end{figure}
%%%%%%%%%%%%%%%%%%%%%%%%%%%%%%%%%%%%%%%%%%%%%%%%%%%%%%%%%%%%%%%%%%%%%%%%%%%%%%%%%%%%%%%%%%

We have constructed an alternative ensemble of $10^5$ spectra $\{S_T^n\}$ with substructures closer
to the reference spectrum $S_T$ as follows: at each of $10^5$ extractions, we have randomly chosen a
subset of only $N'_d=10^2$ nuclear input uncertainties (out of $N_d=17,719$) to be varied.
Figure~\ref{Fig_13} is analogous to Fig.~\ref{Fig_11} but shows the 
envelope of such new spectra, which is narrower and closer to $S_T$ by construction.
 Also in this case, by ranking 
variant spectra with a $\chi^2_{S,n}$ as in Eq.~(\ref{Penalty}) (with $N_d$ replaced by $N'_d$),
the $N_\sigma$ bands would be only marginally smaller than the envelope, confirming that
substructure amplitudes do not scale with $N_\sigma$. 

\begin{figure}[t]
\begin{minipage}[c]{0.9\textwidth}
\includegraphics[width=.49\textwidth]{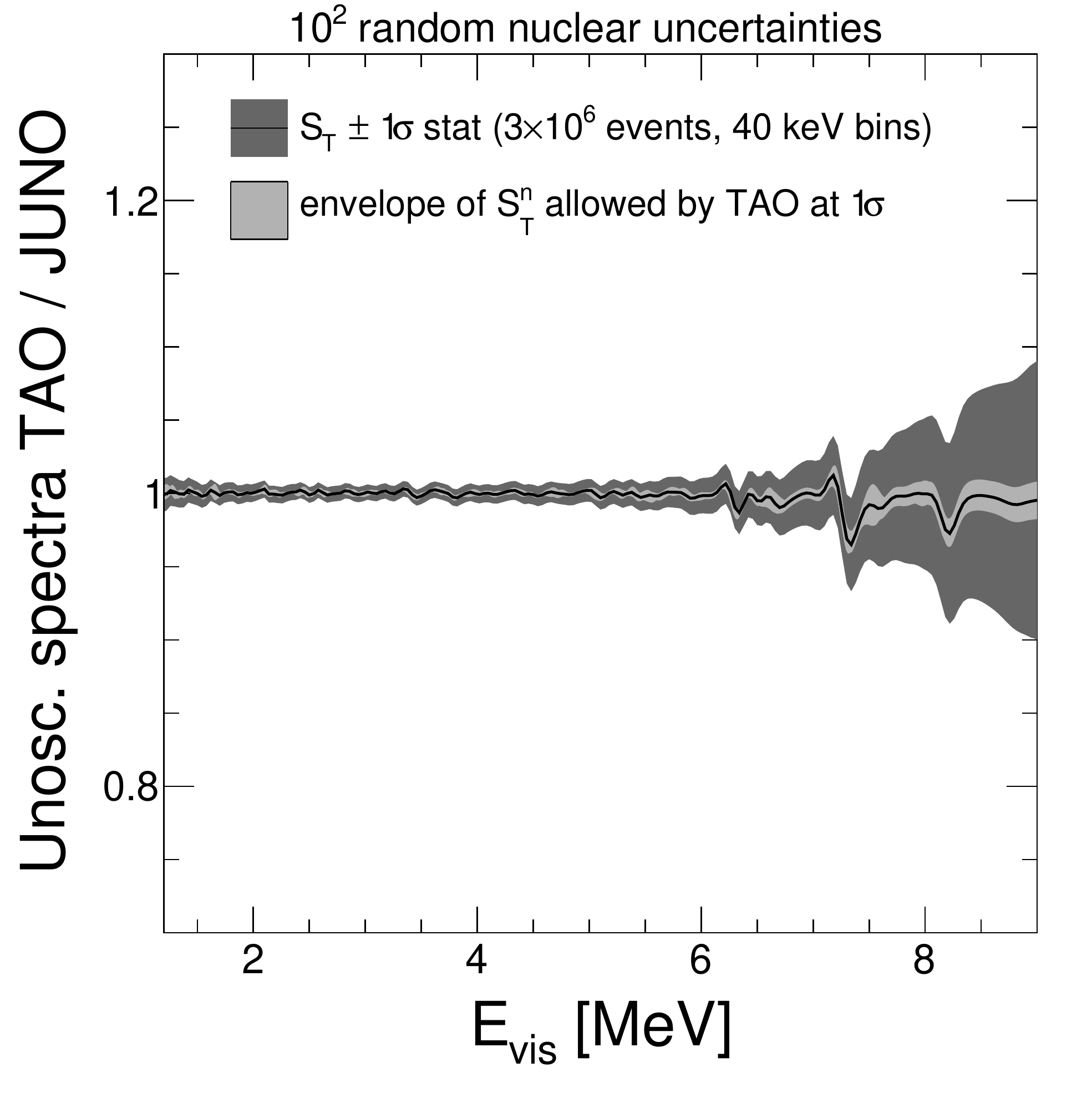}
\caption{\label{Fig_15}
\footnotesize Spectral ensembles in TAO. As in Fig.~\ref{Fig_13}, but with the
light gray  band representing the
envelope of spectra allowed at $1\sigma$ by TAO after collecting $3\times 10^6$ events.
} 
\end{minipage}
\end{figure}

Figure~\ref{Fig_14} is analogous to Fig.~\ref{Fig_12} but with the new ensemble of spectra.
In this case, $O(10^6)$ TAO events are required to start reducing the  fractions
of spectra allowed at $N_\sigma$. For $3\times10^6$ events, using Eq.~(\ref{TAOchi2}),
we find that the envelope of spectra surviving at $1\sigma$ is as shown in Fig.~\ref{Fig_15}. 
The envelopes at $2\sigma$ and $3\sigma$ (not shown) are about a factor of 2 and 3 larger
than the light gray band in Fig.~\ref{Fig_15}, suggesting that the fit to TAO data 
tends to linearize the scaling of allowed substructure amplitudes with $N_\sigma$.

Concerning the JUNO sensitivity to mass ordering, we now find a slight reduction 
of $\Delta \chi^2(\mathrm{IO}-\mathrm{NO})$, amounting to $-0.4$ when scanning over the whole 
new set of $\{S_J^n\}$; this reduction is halved to $-0.2$ when this set is reduced
by TAO via  Eq.~(\ref{TAOchi2}). These relatively small effects, derived through
a $\chi^2$ analysis, agree with the Fourier-analysis findings of \cite{Danielson:2018tzi}: 
variant spectral substructures appear to play a little role in the JUNO sensitivity
to mass ordering, as far as known nuclear uncertainties are concerned.%
%------------------
\footnote{In addition, we find that the fractional precision $\sigma_p/p$ of the parameters
in Fig.~\ref{Fig_10} remains the same, except for a slight reduction by $\sim 20\%$ 
for the $\Delta m^2_{ee}$ uncertainty. The parameter $\Delta m^2_{ee}$ governs the frequency of fast oscillations in JUNO,
and is thus more subject to ``noisy'' substructure variations.}
%------------------- 
 The role is even more marginal with the help of TAO. Of course,
if all substructures shapes were hypothetically allowed, including oscillatory ones appropriately tuned to
``undo'' the IO--NO probability differences, then the sensitivity reduction 
would be higher \cite{Forero:2017vrg}, at the price of introducing ad hoc ``unknown'' errors,  
not belonging to those parametrized in nuclear databases.

%%%%%%%%%%%%%%%%%%%%%%%%%%%%%%%%%%%%%%%%%%%%%%%%%%%%%%%%%%%%%%%%%%%%%%%%%%%%%%%%%%%%%%%%%%
%%%%%%%%%%%%%%%%%%%%%%%%%%%%%%%%%%%%%%%%%%%%%%%%%%%%%%%%%%%%%%%%%%%%%%%%%%%%%%%%%%%%%%%%%%

From the exercises in this Subsection and in the previous one we learn that, once 
TAO spectral data and an associated reference model spectrum $S_T(E_\mathrm{vis})$
will be available, it will be important to sample very densely the functional neighborhood
of such spectrum, in order to study the residual effects of allowed variant spectra in JUNO.
 Brute force variations of all  the $O(10^5)$ nuclear decay parameters may lead to undersampling issues
in this context. Reduction to a limited number of nuclear error sources appears to be
a better strategy. While we have arbitrarily limited this number to $10^2$ random
error sources, future studies may motivate on more physical grounds
a limited subset of nuclear errors (plus possible covariances), related to the decays producing the most pronounced substructures
in TAO. If the nuclear physics of reactor neutrino spectra will not be well understood even in 
in the TAO era, these ``known'' error sources may be cautiously supplemented (but not replaced) by 
some extra errors for ``unknown'' substructures.

%%%%%%%%%%%%%%%%%%%%%%%%%%%%%%%%%%%%%%%%%%%%%%%
\section{Summary and conclusions}
\label{Sec:Final}
%%%%%%%%%%%%%%%%%%%%%%%%%%%%%%%%%%%%%%%%%%%%%%%

The next-generation, medium baseline reactor neutrino experiment JUNO (in construction) is planned to probe the full pattern of
$\overline\nu_e$ disappearance for $L/E\sim (\mathrm{few\ MeV})/(53\ \mathrm{km})$, including the precision measurements of oscillations induced by the $(\delta m^2,\,\theta_12)$ and
$(\Delta m^2,\,\theta_{13})$ mass-mixing pairs, and their interference effects governed by the
neutrino mass ordering, namely sign$(\delta m^2/\Delta m^2)$. The supplementary detector TAO is expected to
monitor the unoscillated flux close to one reactor core, with about a factor $\times 2$ improvement in energy resolution
and with $\times 30$ more events than in JUNO.

In this work we have studied the relations between the observable event spectra in TAO ($S_T$) and JUNO ($S_J$),
in the simplifying assumption that they are generated by the same unobservable neutrino spectrum ($S_\nu$),
including fine-structure features as emerging in summation calculations. We have used
the publicly available Oklo toolkit \cite{Oklo} to generate a  reference spectrum $S_\nu$, as well as a number of variants $S^n_\nu$ corresponding to changes in the (thousands) of nuclear inputs describing fission yields, branching ratios
and endpoint energies. Our methodology can be applied to more updated nuclear databases, which are currently 
being developed  and endowed with preliminary error covariances (not included in this work).

After reviewing
in detail the different and non-negligible effects 
of energy resolution and nucleon recoil on the observable spectra, we have shown that
a model spectrum $S_T$ at TAO site can be mapped into a corresponding spectrum $S_J$ at JUNO via 
well-defined convolutions,
without using the (more detailed) information contained in the source neutrino spectrum $S_\nu$. The mapping
$S_T\to S_J$ is exact in the hypothetical case of no oscillations, and can be generalized with excellent accuracy
to the real case  with oscillations, via an ansatz on the effective disappearance probability. 
The prospective $\chi^2$ analysis of JUNO data confirms the validity of the mapping, and allows to discuss
the impact of uncertainties related to oscillation parameters, energy-scale and flux-shape  systematics.

We have also analyzed the effect of known nuclear input uncertainties, by generating bundles of variant spectra 
with the Oklo toolkit. We highlight several statistical issues arising from sampling a large number of variable inputs. 
We find that the bundles must densely sample the neighbourhood of the reference spectrum, in order to produce
a detectable effect on the JUNO $\chi^2$ function in numerical experiments. In this case (realized by sampling 
only a random subset of nuclear uncertainties), the effect turns
out to be small (in agreement with \cite{Danielson:2018tzi}), and can be further reduced by adding TAO constraints.
These results, based on ``known'' nuclear
inputs, also suggest some cautionary  comments on 
parametrizations of ``unknown'' substructure uncertainties, in terms of variances of binned bundles.

We have argued that, when TAO data will be available, an optimal strategy to deal with small-scale spectral shape
uncertainties will be to focus on a few prominent visible substructures and related
nuisance parameters, in order to build a dense ensemble of TAO spectral variants,  to be mapped in JUNO
and compared with its data. Optimal constructions for such variant ensembles, possibly with covariances 
of known errors and with allowance for
extra unknown errors, as well as for corrections due
to different fuel components in the TAO and JUNO sources,
are left to future studies. We conclude by observing that, after only two decades 
from the discovery of neutrino oscillations, the JUNO and TAO projects are bringing the
discussion of precision oscillometry to an unprecedented level of details, 
whose deeper understanding will require further advances at the junction of neutrino and nuclear physics.

%%%%%%%%%%%%%%%%%%%%%%%%%%%%%%%%%%%%%%%%%%%%%%%%%%%%%%%%%%%%%%%%%%%%%%%%%%%%%%%%%%%%%%%%%%%%%%%%%%%%
%%%%%%%%%%%%%%%%%%%%%%%%%%%%%%%%%%%%%%%%%%%%%%%%%%%%%%%%%%%%%%%%%%%%%%%%%%%%%%%%%%%%%%%%%%%%%%%%%%%%

\acknowledgments

We are grateful to Anna Hayes and Alejandro Sonzogni for useful discussions about fine structures in reactor neutrino spectra.
We thank Jun Cao, Gioacchino Ranucci and Monica Sisti for providing us with early information about the TAO project.  
Preliminary results of this work were shown by A.M.\ at the Conference {\em TAUP 2019\/}  \cite{Marrone:2020oem}.  
This work is partly supported by the Italian Ministero dell'Universit\`a e Ricerca (MUR) through
the research grant no.~2017W4HA7S ``NAT-NET: Neutrino and Astroparticle Theory Network'' under the program PRIN 2017,
and by the Istituto Nazionale di Fisica 
Nucleare (INFN) through the ``Theoretical Astroparticle Physics''  (TAsP) project.
The work of F.C.\ is supported by the Deutsche Forschungsgemeinschaft 
through Grants SFB-1258 ``Neutrinos and Dark Matter in Astro- and Particle Physics (NDM)'' 
and EXC 2094 ``ORIGINS: From the Origin of the Universe to the First Building Blocks of Life''.

%%%%%%%%%%%%%%%%%%%%%%%%%%%%%%%%%%%%%%%%%%%%%%%%%%%%%%%%%%%%%%%%%%%%%%%%%%%%%%%%%%%%%%%%%%%%%%%%


\begin{thebibliography}{999}

%\cite{History}
\bibitem{History}
{\em History of the Neutrino 1930-2018}, Proceedings of the International Conference on History of the Neutrino
(Paris, France, 2018), ed.~by M.~Cribier, J.~Dumarchez and D. Vignaud (AstroParticle and Cosmology Laboratory (APC), Paris, 2019)
[arXiv:1902.03281 [hep-ph]]. Website: \verb|http://neutrinohistory2018.in2p3.fr|

%cite{Bilenky:2019qcm}
\bibitem{Bilenky:2019qcm}
S.~Bilenky,
``Prehistory of neutrino oscillation,''
[arXiv:1902.10052 [physics.hist-ph]].

%\cite{Jarlskog:2019axp}
\bibitem{Jarlskog:2019axp}
C.~Jarlskog,
``Birth of the neutrino, from Pauli to the Reines-Cowan experiment,''
in \cite{History},
[arXiv:1902.03281 [hep-ph]].

%\cite{Vogel:2019fnm}
\bibitem{Vogel:2019fnm}
P.~Vogel,
``Reactor neutrinos: toward oscillations,''
in \cite{History},
[arXiv:1902.03281 [hep-ph]].

%\cite{Lasserre:2019spd}
\bibitem{Lasserre:2019spd}
T.~Lasserre,
``Reactor Neutrinos after CHOOZ and KamLAND,''
in \cite{History}.


%\cite{Cao:2017drk} REVIEW
\bibitem{Cao:2017drk}
L.~J.~Wen, J.~Cao and Y.~Wang,
``Reactor Neutrino Experiments: Present and Future,''
Ann. Rev. Nucl. Part. Sci. \textbf{67}, 183-211 (2017)
doi:10.1146/annurev-nucl-101916-123318
[arXiv:1803.10162 [hep-ex]].


%\cite{Qian:2018wid}
\bibitem{Qian:2018wid}
X.~Qian and J.~Peng,
``Physics with Reactor Neutrinos,''
Rept. Prog. Phys. \textbf{82}, no.3, 036201 (2019)
doi:10.1088/1361-6633/aae881
[arXiv:1801.05386 [hep-ex]].

%\cite{Antonelli:2020uui} REVIEW
\bibitem{Antonelli:2020uui}
V.~Antonelli, L.~Miramonti and G.~Ranucci,
``Present and Future Contributions of Reactor Experiments to Mass Ordering and Neutrino Oscillation Studies,''
Universe \textbf{6}, no.4, 52 (2020)
doi:10.3390/universe6040052

%\cite{Gando:2013nba}
\bibitem{Gando:2013nba}
A.~Gando \textit{et al.} [KamLAND],
``Reactor On-Off Antineutrino Measurement with KamLAND,''
Phys. Rev. D \textbf{88}, no.3, 033001 (2013)
doi:10.1103/PhysRevD.88.033001
[arXiv:1303.4667 [hep-ex]].

\bibitem{Adey:2018zwh} 
  D.~Adey {\it et al.} [Daya Bay Collaboration],
  ``Measurement of the Electron Antineutrino Oscillation with 1958 Days of Operation at Daya Bay,''
  Phys.\ Rev.\ Lett.\  {\bf 121}, no. 24, 241805 (2018)
  doi:10.1103/PhysRevLett.121.241805
  [arXiv:1809.02261 [hep-ex]].

\bibitem{Bak:2018ydk} 
  G.~Bak {\it et al.} [RENO Collaboration],
  ``Measurement of Reactor Antineutrino Oscillation Amplitude and Frequency at RENO,''
  Phys.\ Rev.\ Lett.\  {\bf 121}, no. 20, 201801 (2018)
  doi:10.1103/PhysRevLett.121.201801
  [arXiv:1806.00248 [hep-ex]].


\bibitem{DoubleChooz:2019qbj}
H.~de Kerret \textit{et al.} [Double Chooz],
``First Double Chooz $\mathbf{\theta_{13}}$ Measurement via Total Neutron Capture Detection,''
doi:10.1038/s41567-020-0831-y
[arXiv:1901.09445 [hep-ex]].

\bibitem{PDG} 
M.~C.~Gonzalez-Garcia and M.~Yokoyama, review on ``Neutrino Masses, Mixing and Oscillations'' 
in M.~Tanabashi {\it et al.} [Particle Data Group],
  ``Review of Particle Physics,''
  Phys.\ Rev.\ D {\bf 98}, no. 3, 030001 (2018) and 2019 update,
  doi:10.1103/PhysRevD.98.030001 
  
  
%\cite{Petcov:2001sy}
\bibitem{Petcov:2001sy}
S.~Petcov and M.~Piai,
``The LMA MSW solution of the solar neutrino problem, inverted neutrino mass hierarchy and reactor neutrino experiments,''
Phys. Lett. B \textbf{533}, 94-106 (2002)
doi:10.1016/S0370-2693(02)01591-5
[arXiv:hep-ph/0112074 [hep-ph]].

%\cite{Li:2013zyd}
\bibitem{Li:2013zyd}
Y.~F.~Li, J.~Cao, Y.~Wang and L.~Zhan,
``Unambiguous Determination of the Neutrino Mass Hierarchy Using Reactor Neutrinos,''
Phys. Rev. D \textbf{88}, 013008 (2013)
doi:10.1103/PhysRevD.88.013008
[arXiv:1303.6733 [hep-ex]].


%\cite{An:2015jdp}
\bibitem{An:2015jdp}
F.~An \textit{et al.} [JUNO],
``Neutrino Physics with JUNO,''
J. Phys. G \textbf{43}, no.3, 030401 (2016)
doi:10.1088/0954-3899/43/3/030401
[arXiv:1507.05613 [physics.ins-det]].

%\cite{Hayes:2016qnu}
\bibitem{Hayes:2016qnu}
A.~Hayes and P.~Vogel,
``Reactor Neutrino Spectra,''
Ann. Rev. Nucl. Part. Sci. \textbf{66}, 219-244 (2016)
doi:10.1146/annurev-nucl-102115-044826
[arXiv:1605.02047 [hep-ph]].

%\cite{Huber:2016fkt}
\bibitem{Huber:2016fkt}
P.~Huber,
``Reactor antineutrino fluxes Ð Status and challenges,''
Nucl. Phys. B \textbf{908}, 268-278 (2016)
doi:10.1016/j.nuclphysb.2016.04.012
[arXiv:1602.01499 [hep-ph]].

%\cite{Dentler:2017tkw}
\bibitem{Dentler:2017tkw}
M.~Dentler, A.~Hern\'andez-Cabezudo, J.~Kopp, M.~Maltoni and T.~Schwetz,
``Sterile neutrinos or flux uncertainties? Ñ Status of the reactor anti-neutrino anomaly,''
JHEP \textbf{11}, 099 (2017)
doi:10.1007/JHEP11(2017)099
[arXiv:1709.04294 [hep-ph]].


%\cite{Giunti:2019qlt}
\bibitem{Giunti:2019qlt}
C.~Giunti, Y.~Li, B.~Littlejohn and P.~Surukuchi,
``Diagnosing the Reactor Antineutrino Anomaly with Global Antineutrino Flux Data,''
Phys. Rev. D \textbf{99}, no.7, 073005 (2019)
doi:10.1103/PhysRevD.99.073005
[arXiv:1901.01807 [hep-ph]].


%\cite{Berryman:2020agd}
\bibitem{Berryman:2020agd}
J.~M.~Berryman and P.~Huber,
``Sterile Neutrinos and the Global Reactor Antineutrino Dataset,''
[arXiv:2005.01756 [hep-ph]].



%\cite{Zacek:2018bij}
\bibitem{Zacek:2018bij}
V.~Zacek, G.~Zacek, P.~Vogel and J.~Vuilleumier,
``Evidence for a 5 MeV Spectral Deviation in the Goesgen Reactor Neutrino Oscillation Experiment,''
[arXiv:1807.01810 [hep-ex]].

%\cite{Ko:2016owz}
\bibitem{Ko:2016owz}
Y.~Ko \textit{et al.} [NEOS],
``Sterile Neutrino Search at the NEOS Experiment,''
Phys. Rev. Lett. \textbf{118}, no.12, 121802 (2017)
doi:10.1103/PhysRevLett.118.121802
[arXiv:1610.05134 [hep-ex]].



%\cite{Huber:2016xis} FUELS
\bibitem{Huber:2016xis}
P.~Huber,
``NEOS Data and the Origin of the 5 MeV Bump in the Reactor Antineutrino Spectrum,''
Phys. Rev. Lett. \textbf{118}, no.4, 042502 (2017)
doi:10.1103/PhysRevLett.118.042502
[arXiv:1609.03910 [hep-ph]].


%\cite{Gebre:2017vmm}  FUELS
\bibitem{Gebre:2017vmm}
Y.~Gebre, B.~Littlejohn and P.~Surukuchi, 
``Prospects for Improved Understanding of Isotopic Reactor Antineutrino Fluxes,''
Phys. Rev. D \textbf{97}, no.1, 013003 (2018)
doi:10.1103/PhysRevD.97.013003
[arXiv:1709.10051 [hep-ph]].

%\cite{Giunti:2017nww} FUELS
\bibitem{Giunti:2017nww}
C.~Giunti,
``Improved Determination of the $^{235}\text{U}$ and $^{239}\text{Pu}$ Reactor Antineutrino Cross Sections per Fission,''
Phys. Rev. D \textbf{96}, no.3, 033005 (2017)
doi:10.1103/PhysRevD.96.033005
[arXiv:1704.02276 [hep-ph]].

%\cite{Hayes:2017res} FUELS
\bibitem{Hayes:2017res}
A.~C.~Hayes, G.~Jungman, E.~McCutchan, A.~A.~Sonzogni, G.~T.~Garvey and X.~Wang,
``Analysis of the Daya Bay Reactor Antineutrino Flux Changes with Fuel Burnup,''
Phys. Rev. Lett. \textbf{120}, no.2, 022503 (2018)
doi:10.1103/PhysRevLett.120.022503
[arXiv:1707.07728 [nucl-th]].


%\cite{Adey:2018qct}
\bibitem{Adey:2018qct}
D.~Adey \textit{et al.} [Daya Bay],
``Improved Measurement of the Reactor Antineutrino Flux at Daya Bay,''
Phys. Rev. D \textbf{100}, no.5, 052004 (2019)
doi:10.1103/PhysRevD.100.052004
[arXiv:1808.10836 [hep-ex]].


%\citecAdey:2019ywk}
\bibitem{Adey:2019ywk}
D.~Adey \textit{et al.} [Daya Bay],
``Extraction of the $^{235}$U and $^{239}$Pu Antineutrino Spectra at Daya Bay,''
Phys. Rev. Lett. \textbf{123}, no.11, 111801 (2019)
doi:10.1103/PhysRevLett.123.111801
[arXiv:1904.07812 [hep-ex]].


%\cite{RENO:2018pwo}
\bibitem{RENO:2018pwo}
G.~Bak \textit{et al.} [RENO],
``Fuel-composition dependent reactor antineutrino yield at RENO,''
Phys. Rev. Lett. \textbf{122}, no.23, 232501 (2019)
doi:10.1103/PhysRevLett.122.232501
[arXiv:1806.00574 [hep-ex]].


%\cite{Ashenfelter:2018jrx}
\bibitem{Ashenfelter:2018jrx}
J.~Ashenfelter \textit{et al.} [PROSPECT],
``Measurement of the Antineutrino Spectrum from $^{235}$U Fission at HFIR with PROSPECT,''
Phys. Rev. Lett. \textbf{122}, no.25, 251801 (2019)
doi:10.1103/PhysRevLett.122.251801
[arXiv:1812.10877 [nucl-ex]].

%\cite{Hayes:2015yka} DATABASE
\bibitem{Hayes:2015yka}
A.~Hayes, J.~Friar, G.~Garvey, D.~Ibeling, G.~Jungman, T.~Kawano and R.~W.~Mills,
``Possible origins and implications of the shoulder in reactor neutrino spectra,''
Phys. Rev. D \textbf{92}, no.3, 033015 (2015)
doi:10.1103/PhysRevD.92.033015
[arXiv:1506.00583 [nucl-th]].

%\cite{Sonzogni:2016yac} DATABASE
\bibitem{Sonzogni:2016yac}
A.~A.~Sonzogni, E.~A.~McCutchan, T.~D.~Johnson and P.~Dimitriou,
``Effects of Fission Yield Data in the Calculation of Antineutrino Spectra for U235(n,fission) at Thermal and Fast Neutron Energies,''
Phys. Rev. Lett. \textbf{116}, no.13, 132502 (2016)
doi:10.1103/PhysRevLett.116.132502

%\cite{Sonzogni:2017wxy}
\bibitem{Sonzogni:2017wxy}
A.~Sonzogni, E.~McCutchan and A.~Hayes,
``Dissecting Reactor Antineutrino Flux Calculations,''
Phys. Rev. Lett. \textbf{119}, no.11, 112501 (2017)
doi:10.1103/PhysRevLett.119.112501
%

%\cite{Ma:2018fqf} DATABASES
\bibitem{Ma:2018fqf}
X.~Ma, L.~Yang, L.~Zhan, F.~An and J.~Cao,
``Investigation of antineutrino spectral anomaly with updated nuclear database,''
[arXiv:1807.09265 [nucl-ex]].

%\cite{Mention:2017dyq}
\bibitem{Mention:2017dyq}
G.~Mention, M.~Vivier, J.~Gaffiot, T.~Lasserre, A.~Letourneau and T.~Materna,
``Reactor antineutrino shoulder explained by energy scale nonlinearities?,''
Phys. Lett. B \textbf{773}, 307-312 (2017)
doi:10.1016/j.physletb.2017.08.035
[arXiv:1705.09434 [hep-ex]].

%\cite{Hardy:1977suw}
\bibitem{Hardy:1977suw}
J.~Hardy, L.~Carraz, B.~Jonson and P.~Hansen,
``The essential decay of pandemonium: A demonstration of errors in complex beta-decay schemes,''
Phys. Lett. B \textbf{71}, 307-310 (1977)
doi:10.1016/0370-2693(77)90223-4


%%%+++++++++++++++++++++++++++++++++++++++++++++++++++++++

%\cite{Fallot:2012jv}
\bibitem{Fallot:2012jv}
M.~Fallot {\em et al.},
``New antineutrino energy spectra predictions from the summation of beta decay branches of the fission products,''
Phys. Rev. Lett. \textbf{109}, 202504 (2012)
doi:10.1103/PhysRevLett.109.202504
[arXiv:1208.3877 [nucl-ex]].


%\cite{Rasco:2016leq}
\bibitem{Rasco:2016leq}
B.~Rasco {\em et al.},
``Decays of the Three Top Contributors to the Reactor $\overline \nu_e$ High-Energy Spectrum, $^{92}$Rb, $^{96gs}$Y, and $^{142}$Cs, Studied with Total Absorption Spectroscopy,''
Phys. Rev. Lett. \textbf{117}, no.9, 092501 (2016)
doi:10.1103/PhysRevLett.117.092501


%\cite{Guadilla:2019gws}
\bibitem{Guadilla:2019gws}
V.~Guadilla {\em et al.},
``Total absorption $\gamma$-ray spectroscopy of the $\beta$-delayed neutron emitters $^{137}$I and $^{95}$Rb,''
Phys. Rev. C \textbf{100}, no.4, 044305 (2019)
doi:10.1103/PhysRevC.100.044305
[arXiv:1907.02748 [nucl-ex]].


%\cite{Guadilla:2019zwz}
\bibitem{Guadilla:2019zwz}
V.~Guadilla {\em et al.},
``Total absorption $\gamma$-ray spectroscopy of niobium isomers,''
Phys. Rev. C \textbf{100}, no.2, 024311 (2019)
doi:10.1103/PhysRevC.100.024311
[arXiv:1904.07036 [nucl-ex]].


%\cite{Guadilla:2019aiq}
\bibitem{Guadilla:2019aiq}
V.~Guadilla {\em et al.},
``Large Impact of the Decay of Niobium Isomers on the Reactor ${\overline{{\nu}}}_{e}$ Summation Calculations,''
Phys. Rev. Lett. \textbf{122}, no.4, 042502 (2019)
doi:10.1103/PhysRevLett.122.042502


%\cite{Estienne:2019ujo}
\bibitem{Estienne:2019ujo}
M.~Estienne {\em et al.}, ``Updated Summation Model: An Improved Agreement with the Daya Bay Antineutrino Fluxes,''
Phys. Rev. Lett. \textbf{123}, no.2, 022502 (2019)
doi:10.1103/PhysRevLett.123.022502
[arXiv:1904.09358 [nucl-ex]].

%\cite{Guadilla:2020pjj}
\bibitem{Guadilla:2020pjj}
V.~Guadilla {\em et al.}, 
``Determination of Beta Decay Ground State Feeding of Nuclei of Importance for Reactor Applications,''
[arXiv:2005.08780 [nucl-ex]].


%\cite{Hayes:2013wra}
\bibitem{Hayes:2013wra}
A.~Hayes, J.~Friar, G.~Garvey, G.~Jungman and G.~Jonkmans,
``Systematic Uncertainties in the Analysis of the Reactor Neutrino Anomaly,''
Phys. Rev. Lett. \textbf{112}, 202501 (2014)
doi:10.1103/PhysRevLett.112.202501
[arXiv:1309.4146 [nucl-th]].

%\cite{Wang:2017htp}
\bibitem{Wang:2017htp}
X.~Wang and A.~Hayes,
``Weak magnetism correction to allowed $\beta$ decay for reactor antineutrino spectra,''
Phys. Rev. C \textbf{95}, no.6, 064313 (2017)
doi:10.1103/PhysRevC.95.064313
[arXiv:1702.07520 [nucl-th]].



%\cite{Li:2019quv}
\bibitem{Li:2019quv}
Y.~Li and D.~Zhang,
``New Realization of the Conversion Calculation for Reactor Antineutrino Fluxes,''
Phys. Rev. D \textbf{100}, no.5, 053005 (2019)
doi:10.1103/PhysRevD.100.053005
[arXiv:1904.07791 [hep-ph]].


%\cite{Fang:2015cma}
\bibitem{Fang:2015cma}
D.~L.~Fang and B.~Brown,
``Effect of first forbidden decays on the shape of neutrino spectra,''
Phys. Rev. C \textbf{91}, no.2, 025503 (2015)
doi:10.1103/PhysRevC.93.049903
[arXiv:1502.02246 [nucl-th]].


%\cite{Fang:2020emq}
\bibitem{Fang:2020emq}
D.~Fang, Y.~Li and D.~Zhang,
``Ab initio calculations of reactor antineutrino fluxes with exact lepton wave functions,''
[arXiv:2001.01689 [hep-ph]].

%\cite{Yoshida:2018zga}
\bibitem{Yoshida:2018zga}
T.~Yoshida, T.~Tachibana, S.~Okumura and S.~Chiba,
``Spectral anomaly of reactor antineutrinos based on theoretical energy spectra,''
Phys. Rev. C \textbf{98}, no.4, 041303 (2018)
doi:10.1103/PhysRevC.98.041303



%\cite{Petkovic:2019wyw}
\bibitem{Petkovic:2019wyw}
J.~Petkovi\'c, T.~Marketin, G.~Mart\'inez-Pinedo and N.~Paar,
``Self-consistent calculation of the reactor antineutrino spectra including forbidden transitions,''
J. Phys. G \textbf{46}, no.8, 085103 (2019)
doi:10.1088/1361-6471/ab28f5
[arXiv:1903.06192 [nucl-th]].



%\cite{Hayen:2018uyg}
\bibitem{Hayen:2018uyg}
L.~Hayen, J.~Kostensalo, N.~Severijns and J.~Suhonen,
``First forbidden transitions in the reactor anomaly,''
Phys. Rev. C \textbf{100}, no.5, 054323 (2019)
doi:10.1103/PhysRevC.100.054323
[arXiv:1908.08302 [nucl-th]].


%\cite{Hayen:2019ieh}
\bibitem{Hayen:2019ieh}
L.~Hayen, J.~Kostensalo, N.~Severijns and J.~Suhonen,
``First-forbidden transitions in reactor antineutrino spectra,''
Phys. Rev. C \textbf{99}, no.3, 031301 (2019)
doi:10.1103/PhysRevC.99.031301







%%%+++++++++++++++++++++++++++++++++++++++++++++++++++++++

%\cite{Dwyer:2014eka}
\bibitem{Dwyer:2014eka}
D.~Dwyer and T.~Langford,
``Spectral Structure of Electron Antineutrinos from Nuclear Reactors,''
Phys. Rev. Lett. \textbf{114}, no.1, 012502 (2015)
doi:10.1103/PhysRevLett.114.012502
[arXiv:1407.1281 [nucl-ex]].


%\cite{Sonzogni:2015aoa}
\bibitem{Sonzogni:2015aoa}
A.~Sonzogni, T.~Johnson and E.~McCutchan,
``Nuclear structure insights into reactor antineutrino spectra,''
Phys. Rev. C \textbf{91}, no.1, 011301 (2015)
doi:10.1103/PhysRevC.91.011301


%\cite{Sonzogni:2017voo}
\bibitem{Sonzogni:2017voo}
A.~Sonzogni, M.~Nino and E.~McCutchan,
``Revealing Fine Structure in the Antineutrino Spectra From a Nuclear Reactor,''
Phys. Rev. C \textbf{98}, no.1, 014323 (2018)
doi:10.1103/PhysRevC.98.014323
[arXiv:1710.00092 [nucl-th]].



%\cite{Forero:2017vrg}
\bibitem{Forero:2017vrg}
D.~V.~Forero, R.~Hawkins and P.~Huber,
``The benefits of a near detector for JUNO,''
[arXiv:1710.07378 [hep-ph]].


%\cite{Danielson:2018tzi} HAYES WORK
\bibitem{Danielson:2018tzi}
D.~Danielson, A.~Hayes and G.~Garvey,
``Reactor Neutrino Spectral Distortions Play Little Role in Mass Hierarchy Experiments,''
Phys. Rev. D \textbf{99}, no.3, 036001 (2019)
doi:10.1103/PhysRevD.99.036001
[arXiv:1808.03276 [hep-ph]].


%\cite{Cheng:2020ivh}
\bibitem{Cheng:2020ivh}
Z.~Cheng, N.~Raper, W.~Wang, C.~F.~Wong and J.~Zhang,
``Potential impact of sub-structure on the resolution of neutrino mass hierarchy at medium-baseline reactor neutrino oscillation experiments,''
[arXiv:2004.11659 [hep-ex]].


%\cite{IAEA-Vienna}
\bibitem{IAEA-Vienna}
Technical Meeting on Nuclear Data for Anti-neutrino Spectra and Their Applications (IAEA, Vienna, 2019). Contributions available at
the website: \verb|https://www-nds.iaea.org/index-meeting-crp/Antineutrinos|

%\cite{Mikaelyan:1998yg}
\bibitem{Mikaelyan:1998yg}
L.~Mikaelyan and V.~Sinev,
``Searches for sterile neutrinos at reactor,''
Phys. Atom. Nucl. \textbf{62}, 2008-2012 (1999)
[arXiv:hep-ph/9811228 [hep-ph]].

%\cite{Mikaelyan:1999pm}
\bibitem{Mikaelyan:1999pm}
L.~Mikaelyan and V.~Sinev,
%``Neutrino oscillations at reactors: What next?,''
Phys. Atom. Nucl. \textbf{63}, 1002-1006 (2000)
doi:10.1134/1.855739
[arXiv:hep-ex/9908047 [hep-ex]].

%\cite{Wang-2017}
\bibitem{Wang-2017}
Y.~Wang [JUNO], 
``Unknowns of Neutrinos,''
in the final discussion on ``Perspectives in Neutrino and Multi Messenger Physics'' 
at {\em NEUTEL 2017}, XVII International Workshop on Neutrino Telescopes (Venice, Italy, 2017),
available at \verb|https://agenda.infn.it/event/11857|

%\cite{Zhan-2018} KNOWN VS UNKNOWN FINE STRUCTURE UNCERTAINTIES
\bibitem{Zhan-2018}
L.~Zhan [JUNO],
``A High Energy Resolution Detector for the Measurement of Reactor Antineutrino Spectrum''
talk at {\em ESCAPE 2018}, Workshop on Energy Scale Calibration in Antineutrino Precision Experiments 
(Heidelberg, Germany, 2018), available at
\verb|https://www.mpi-hd.mpg.de/escape2018|,
doi:10.5281/zenodo.1314423

%\cite{Wonsak-2018}
\bibitem{Wonsak-2018}
B. Wonsak [JUNO], 
``Status and Prospects of the JUNO Experiment,'' 
talk at {\em Neutrino 2018}, XXVIII International Conference on Neutrino Physics and Astrophysics,  
(Heidelberg, Germany, 2018), available at
\verb|https://www.mpi-hd.mpg.de/nu2018|,
doi:10.5281/zenodo.1286850


%\cite{Cao-2019}
\bibitem{Cao-2019}
Jun Cao [JUNO], 
``Measuring High Resolution Reactor Neutrino Spectrum with JUNO-TAO,'' 
talk at \cite{IAEA-Vienna};
``Reactor neutrino anomalies,''
talk at the {\em 19th Lomonosov Conference on Elementary Particle Physics\/} (Moscow, Russia, 2019),
available at \verb|http://lomcon.ru/?page\_id=204|

%\cite{Ranucci:2018erv}
\bibitem{Ranucci:2018erv}
G.~Ranucci [JUNO],
``JUNO Oscillation Physics Program,''
in the Proceedings of {\em NOW 2018}, Neutrino Oscillation Workshop 2018 (Ostuni, Italy, 2018),
ed.\ by A.~Marrone, A.~Mirizzi and D.~Montanino,
PoS \textbf{NOW2018}, 024 (2018)
doi:10.22323/1.337.0024


%\cite{Wang-2019}
\bibitem{Wang-2019}
W.~Wang [JUNO],
``Taishan Antineutrino Observatory,''
talk at {\em NNN~2019},
20th International Workshop on Next generation Nucleon decay and Neutrino detectors  
(Medellin, Colombia, 2019), available at
\verb|https://indico.cern.ch/event/835190|

%\cite{Sisti-2019}
\bibitem{Sisti-2019}
M.~Sisti [JUNO],
``Physics prospects of the JUNO experiment,''
in the Proceedings of {\em TAUP 2019}, 16th International Conference on Topics in Astroparticle and Underground Physics 
(Toyama, Japan, 2019), ed.~by M.~Nakahata and M.~Ohashi,
J. Phys. Conf. Ser. \textbf{1468}, no.1, 012150 (2020)
doi:10.1088/1742-6596/1468/1/012150

%\cite{Smirnov-2020}
\bibitem{Smirnov-2020}
M.~V.~Smirnov [JUNO],
``Status and physics of the JUNO experiment,''
seminar at the High Energy Physics Division of the Petersburg Nuclear Physics Institute (PNPI, Saint-Petersburg, Russia, January 2020), 
available at \verb|http://hepd.pnpi.spb.ru/hepd/events/seminar_index.html|


\bibitem{TAO-CDR}
A.~Abusleme {\em et al.} [JUNO],
``TAO Conceptual Design Report: A Precision Measurement of the Reactor Antineutrino Spectrum with Sub-percent Energy Resolution,''
[arXiv:2005.08745 [physics.ins-det]].

%\cite{Brown:2018jhj}
\bibitem{Brown:2018jhj}
D.~Brown {\em et al.},
``ENDF/B-VIII.0: The 8th Major Release of the Nuclear Reaction Data Library with CIELO-project Cross Sections, New Standards and Thermal Scattering Data,''
Nucl. Data Sheets \textbf{148}, 1-142 (2018)
doi:10.1016/j.nds.2018.02.001


%\cite{Schmidt:2018hwz}
\bibitem{Schmidt:2018hwz}
K.~Schmidt and B.~Jurado,
``Review on the progress in nuclear fission--experimental methods and theoretical descriptions,''
Rept. Prog. Phys. \textbf{81}, no.10, 106301 (2018)
doi:10.1088/1361-6633/aacfa7
[arXiv:1804.10421 [nucl-th]].


%\cite{Bernstein:2019nqq}
\bibitem{Bernstein:2019nqq}
L.~A.~Bernstein, D.~A.~Brown, A.~J.~Koning, B.~T.~Rearden, C.~E.~Romano, A.~A.~Sonzogni, A.~S.~Voyles and W.~Younes,
``Our Future Nuclear Data Needs,''
Ann. Rev. Nucl. Part. Sci. \textbf{69}, 109-136 (2019)
doi:10.1146/annurev-nucl-101918-023708


%\cite{Ejiri:2019ezh}
\bibitem{Ejiri:2019ezh}
H.~Ejiri, J.~Suhonen and K.~Zuber,
``Neutrino-nuclear responses for astro-neutrinos, single beta decays and double beta decays,''
Phys. Rept. \textbf{797}, 1-102 (2019)
doi:10.1016/j.physrep.2018.12.001


%\cite{Capozzi:2013psa}
\bibitem{Capozzi:2013psa}
F.~Capozzi, E.~Lisi and A.~Marrone,
``Neutrino mass hierarchy and electron neutrino oscillation parameters with one hundred thousand reactor events,''
Phys. Rev. D \textbf{89}, no.1, 013001 (2014)
doi:10.1103/PhysRevD.89.013001
[arXiv:1309.1638 [hep-ph]].


%\cite{Capozzi:2015bpa}
\bibitem{Capozzi:2015bpa}
F.~Capozzi, E.~Lisi and A.~Marrone,
``Neutrino mass hierarchy and precision physics with medium-baseline reactors: Impact of energy-scale and flux-shape uncertainties,''
Phys. Rev. D \textbf{92}, no.9, 093011 (2015)
doi:10.1103/PhysRevD.92.093011
[arXiv:1508.01392 [hep-ph]].


%\cite{Oklo}
\bibitem{Oklo}
D.~Dwyer, ``OKLO: A toolkit for modeling nuclides and
nuclear reactions,'' \verb|https://github.com/dadwyer/oklo| (2015).

%\cite{Littlejohn:2018hqm}  oklo used herein
\bibitem{Littlejohn:2018hqm}
B.~Littlejohn, A.~Conant, D.~Dwyer, A.~Erickson, I.~Gustafson and K.~Hermanek,
``Impact of Fission Neutron Energies on Reactor Antineutrino Spectra,''
Phys. Rev. D \textbf{97}, no.7, 073007 (2018)
doi:10.1103/PhysRevD.97.073007
[arXiv:1803.01787 [nucl-th]].


%\cite{Ciuffoli:2019nli} TAO and FUEL COMPOSITION
\bibitem{Ciuffoli:2019nli}
E.~Ciuffoli, J.~Evslin and H.~Mohammed,
``Uncertainty in the Reactor Neutrino Spectrum and Mass Hierarchy Determination,''
JHEP \textbf{10}, 143 (2019)
doi:10.1007/JHEP10(2019)143
[arXiv:1907.02309 [hep-ph]].
%1 citations counted in INSPIRE as of 28 Apr 2020



%\cite{Vogel:1999zy}
\bibitem{Vogel:1999zy}
P.~Vogel and J.~F.~Beacom,
``Angular distribution of neutron inverse beta decay, $\overline\nu_e+p\to e^+ +n$,''
Phys. Rev. D \textbf{60}, 053003 (1999)
doi:10.1103/PhysRevD.60.053003
[arXiv:hep-ph/9903554 [hep-ph]].
%542 citations counted in INSPIRE as of 30 Apr 2020

%\cite{Strumia:2003zx}
\bibitem{Strumia:2003zx}
A.~Strumia and F.~Vissani,
``Precise quasielastic neutrino/nucleon cross-section,''
Phys. Lett. B \textbf{564}, 42-54 (2003)
doi:10.1016/S0370-2693(03)00616-6
[arXiv:astro-ph/0302055 [astro-ph]].
%308 citations counted in INSPIRE as of 30 Apr 2020


%\cite{Vissani:2014doa} JACOBIAN
\bibitem{Vissani:2014doa}
F.~Vissani,
``Comparative analysis of SN1987A antineutrino fluence,''
J. Phys. G \textbf{42}, 013001 (2015)
doi:10.1088/0954-3899/42/1/013001
[arXiv:1409.4710 [astro-ph.HE]].
%40 citations counted in INSPIRE as of 30 Apr 2020




%\cite{Wei:2020yfs}
\bibitem{Wei:2020yfs}
L.~Wei, L.~Zhan, J.~Cao and W.~Wang,
``Improving the Energy Resolution of the Reactor Antineutrino Energy Reconstruction with Positron Direction,''
[arXiv:2005.05034 [physics.ins-det]].


%\cite{Minakata:2007tn}
\bibitem{Minakata:2007tn}
H.~Minakata, H.~Nunokawa, S.~J.~Parke and R.~Zukanovich Funchal,
``Determination of the Neutrino Mass Hierarchy via the Phase of the Disappearance Oscillation Probability with a Monochromatic $\bar{\nu}_e$ Source,''
Phys. Rev. D \textbf{76}, 053004 (2007)
doi:10.1103/PhysRevD.76.079901
[arXiv:hep-ph/0701151 [hep-ph]].

%\cite{Li:2016txk}
\bibitem{Li:2016txk}
Y.~F.~Li, Y.~Wang and Z.~z.~Xing,
``Terrestrial matter effects on reactor antineutrino oscillations at JUNO or RENO-50: how small is small?,''
Chin. Phys. C \textbf{40}, no.9, 091001 (2016)
doi:10.1088/1674-1137/40/9/091001
[arXiv:1605.00900 [hep-ph]].

%\cite{Khan:2019doq}
\bibitem{Khan:2019doq}
A.~N.~Khan, H.~Nunokawa and S.~J.~Parke,
``Why matter effects matter for JUNO,''
Phys. Lett. B \textbf{803}, 135354 (2020)
doi:10.1016/j.physletb.2020.135354
[arXiv:1910.12900 [hep-ph]].


\bibitem{Capozzi:2018ubv} 
  F.~Capozzi, E.~Lisi, A.~Marrone and A.~Palazzo,
  ``Current unknowns in the three neutrino framework,''
  Prog.\ Part.\ Nucl.\ Phys.\  {\bf 102}, 48 (2018)
  doi:10.1016/j.ppnp.2018.05.005
  [arXiv:1804.09678 [hep-ph]].
 

%\cite{Parke:2016joa}
\bibitem{Parke:2016joa}
S.~Parke,
``What is $\Delta m^2_{ee}$ ?,''
Phys. Rev. D \textbf{93}, no.5, 053008 (2016)
doi:10.1103/PhysRevD.93.053008
[arXiv:1601.07464 [hep-ph]].

%\cite{Qian:2012xh}
\bibitem{Qian:2012xh}
X.~Qian, D.~Dwyer, R.~McKeown, P.~Vogel, W.~Wang and C.~Zhang,
``Mass Hierarchy Resolution in Reactor Anti-neutrino Experiments: Parameter Degeneracies and Detector Energy Response,''
Phys. Rev. D \textbf{87}, no.3, 033005 (2013)
doi:10.1103/PhysRevD.87.033005
[arXiv:1208.1551 [physics.ins-det]].




%\cite{Adey:2019zfo}
\bibitem{Adey:2019zfo}
D.~Adey \textit{et al.} [Daya Bay],
``A high precision calibration of the nonlinear energy response at Daya Bay,''
Nucl. Instrum. Meth. A \textbf{940}, 230-242 (2019)
doi:10.1016/j.nima.2019.06.031
[arXiv:1902.08241 [physics.ins-det]].



%\cite{King:1958zz}
\bibitem{King:1958zz}
R.~King and J.~Perkins,
``Inverse Beta Decay and the Two-Component Neutrino,''
Phys. Rev. \textbf{112}, 963-966 (1958)
doi:10.1103/PhysRev.112.963
%9 citations counted in INSPIRE as of 30 Apr 2020

%\cite{Avignone}
\bibitem{Avignone}
F.~T.~Avignone, S.~M.~Blakenship and C.~W.~Darden,
``Theoretical Fission-Antineutrino Spectrum and Cross Section of the Reaction 
$^{3}\mathrm{He}({\overline{\ensuremath{\nu}}}_{e},{e}^{+})^{3}\mathrm{H}$,''
Phys. Rev. \textbf{170}, 931-934 (1968)
doi:10.1103/PhysRev.170.931
%%%%%%%%%%%%%%

%\cite{Davis:1979gg}
\bibitem{Davis:1979gg}
B.~Davis, P.~Vogel, F.~Mann and R.~Schenter,
``Reactor anti-neutrino spectra and their application to anti-neutrino induced reactions,''
Phys. Rev. C \textbf{19}, 2259-2266 (1979)
doi:10.1103/PhysRevC.19.2259
%97 citations counted in INSPIRE as of 30 Apr 2020


%\cite{Avignone:1980qg}
\bibitem{Avignone:1980qg}
F.~Avignone, III and C.~Greenwood,
``Calculated spectra of antineutrinos from U-235, U-238, and Pu-239, and antineutrino-induced reactions,''
Phys. Rev. C \textbf{22}, 594-605 (1980)
doi:10.1103/PhysRevC.22.594
%49 citations counted in INSPIRE as of 30 Apr 2020

\bibitem{Sonzogni-IAEA}
A.\ Sonzogni, 
``Uncertainty quantification in the summation method for nuclear reactor antineutrinos,'' 
talk at \cite{IAEA-Vienna}.

\bibitem{Sonzogni-AAP}
A.\ Sonzogni, 
``Development of realistic uncertainties in the summation method for nuclear reactor antineutrino applications,''
talk at {\em AAP 2019}, Workshop on Applied Antineutrino Physics (Sun Yat-sen, China, 2019).
Website: \verb|https://indico.cern.ch/event/833568|


\bibitem{Cowan-PDG}
G.\ Cowan, review on ``Probability,'' in 
M. Tanabashi {\em et al.} (Particle Data Group), Phys.\ Rev.\ D {\bf 98}, 030001 (2018) and 2019 update.
Website: \verb|http://pdg.lbl.gov|


%\cite{Ge:2012wj}
\bibitem{Ge:2012wj}
S.~F.~Ge, K.~Hagiwara, N.~Okamura and Y.~Takaesu,
``Determination of mass hierarchy with medium baseline reactor neutrino experiments,''
JHEP \textbf{05}, 131 (2013)
doi:10.1007/JHEP05(2013)131
[arXiv:1210.8141 [hep-ph]].


%\cite{Marrone:2020oem}
\bibitem{Marrone:2020oem}
A.~Marrone, F.~Capozzi and E.~Lisi,
``Impact of theoretical reactor flux uncertainties and of the near detector on the JUNO measurements,''
in the Proceedings of {\em TAUP 2019}, 16th International Conference on Topics in Astroparticle and Underground Physics 
(Toyama, Japan, 2019), ed.~by M.~Nakahata and M.~Ohashi,
J. Phys. Conf. Ser. \textbf{1468}, no.1, 012202 (2020)
doi:10.1088/1742-6596/1468/1/012202





































































  
 





\end{thebibliography}
\end{document}